\documentclass[10pt,journal]{IEEEtran}
\usepackage{cite}
\usepackage{amsmath,amssymb,amsfonts}
\usepackage{algorithmic}
\usepackage{graphicx}
\usepackage{algorithm,algorithmic}
\usepackage{hyperref}
\hypersetup{hidelinks=true}
\usepackage{textcomp}
\usepackage{breqn,xspace}
\usepackage{algorithmic}
\usepackage{color}
\usepackage{subfig}
\usepackage{bm}

\ifodd 0
\newcommand{\rev}[1]{{\color{blue}#1}} 
\else
\newcommand{\rev}[1]{#1}
\fi

\def\BibTeX{{\rm B\kern-.05em{\sc i\kern-.025em b}\kern-.08em
    T\kern-.1667em\lower.7ex\hbox{E}\kern-.125emX}}
\newcommand{\name}{ISAC-Fi\xspace}

\begin{document}

\title{ISAC-Fi: Enabling Full-fledged Monostatic Sensing over Wi-Fi Communication}

\author{Zhe~Chen, Chao~Hu, Tianyue~Zheng, Hangcheng~Cao, Yanbing~Yang, Yen~Chu,\\ Hongbo~Jiang,~\IEEEmembership{Senier Member,~IEEE}, and~Jun~Luo,~\IEEEmembership{Fellow,~IEEE}
\thanks{Zhe Chen is with Institute of Space Internet and School of Computer Science, Fudan University, China (e-mail: zhechen@fudan.edu.cn).}
\thanks{Chao Hu and Yanbing Yang are with the College of Computer Science, Sichuan University, China (e-mail: huchao484@stu.scu.edu.cn; yangyanbing@scu.edu.cn).}
\thanks{
Tianyue Zheng and Jun Luo are with the School of Computer Science and Engineering, Nanyang Technological University, Singapore (e-mail: \{tianyue002, junluo\}@ntu.edu.sg).
}
\thanks{
Hangcheng Cao and Hongbo Jiang are with the College of Computer Science and Electronics Engineering, Hunan University, China (e-mail: \{hangchengcao, hongbojiang\}@hnu.edu.cn).
}
\thanks{
Yen Chu is with University of Electronic Science and Technology of China, China (e-mail: yenchu@uestc.edu.cn).
}
}

\maketitle

\begin{abstract}
Whereas Wi-Fi communications have been exploited for sensing purpose for over a decade, the \textit{bistatic} or \textit{multistatic} nature of Wi-Fi still poses multiple challenges, hampering real-life deployment of \textit{integrated sensing and communication}~(ISAC) within Wi-Fi framework. 
In this paper, we aim to re-design Wi-Fi so that \textit{monostatic} sensing (mimicking radar) can be achieved over the multistatic communication infrastructure.
Specifically, we propose, design, and implement \name as an ISAC-ready Wi-Fi prototype. We first present a novel self-interference cancellation scheme, in order to extract reflected (radio frequency) signals for sensing purpose in the face of transmissions.
We then subtly revise existing Wi-Fi framework so as to seamlessly operate monostatic sensing under Wi-Fi communication standard. Finally, we offer two ISAC-Fi designs: while a USRP-based one emulates a totally re-designed \name device, another plug-and-play design allows for backward compatibility by attaching an extra module to an arbitrary Wi-Fi device.
We perform extensive experiments to validate the efficacy of \name and also to demonstrate its superiority over existing Wi-Fi sensing proposals. 
\end{abstract}


\section{Introduction} \label{sec:intro}
Although the received signal strength carried by Wi-Fi signaling has been exploited by indoor localization for more than two decades~\cite{RADAR-INFOCOM00,Horus-MobiSys05}, the true \textit{Wi-Fi sensing} (i.e., leveraging Wi-Fi communications) only started a decade ago thanks to the ability of extracting Channel State Information (CSI) from data packets~\cite{CSI-CCR11}. In particular, numerous applications of \textit{device-free} Wi-Fi sensing have been proposed to utilize CSI, notably including localization/tracking~\cite{WiDeo-NSDI15,LiFS-MobiCom16,lin2022tracking,Widar2-MobiSys18,lin2022v2i,mDTrack-MobiCom19}, activity/gesture recognition~\cite{WiVi-SIGCOMM13,WiAG-MobiSys17,qiu2024ifvit,EI-MobiCom18,Widar3-MobiSys19,RF-Net,xiao2021onefi}, vital signs monitoring~\cite{VitalSign-MobiHoc15,Fresnel-UbiComp18}, and object identification/imaging~\cite{Soil-MobiCom19,Material-MobiCom19,lin2023fedsn}. Whereas these applications all bear a promising future, the \textit{bistatic}
nature of Wi-Fi infrastructure has largely hampered the deployment of real-life systems. Essentially, as a Wi-Fi communication session involves at least a pair of physically separated transmitter (Tx) and receiver (Rx), any sensing function piggybacking on this infrastructure is subject to the severe constraints imposed by the \textit{physical separation} of Tx and Rx.

\begin{figure}[t]
	\centering
	\subfloat[Conventional Wi-Fi sensing.]{
		\begin{minipage}[b]{\columnwidth}
			\centering
			\includegraphics[width = 0.88\textwidth]{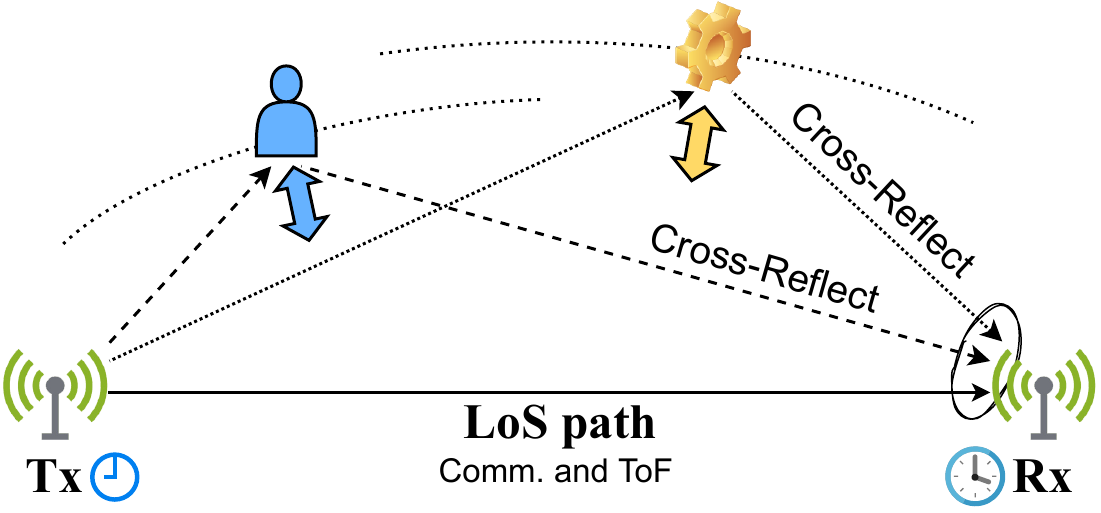}
			\label{sfig:teaser_wifi}
		\end{minipage}
	}
	\\ 
	\vspace{-.5ex}
	\subfloat[ISAC-Fi sensing.]{
		\begin{minipage}[b]{\columnwidth}
			\centering
			\includegraphics[width = 0.88\textwidth]{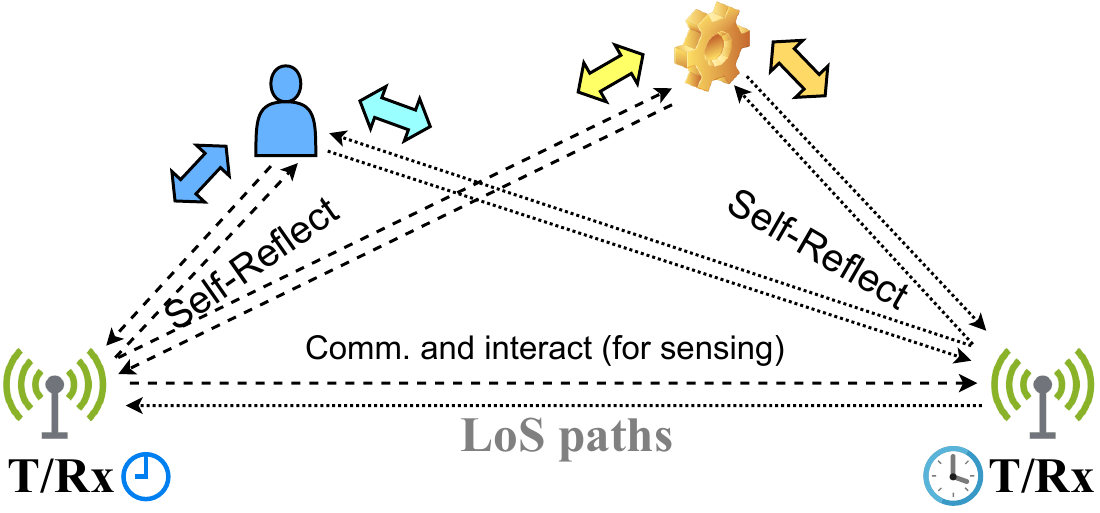}
			\label{sfig:teaser_isacfi}
		\end{minipage}
	}
	\caption{Wi-Fi (a) vs ISAC-Fi (b) sensing. The thin arrows represent RF propagation originated from different Tx's or along distinct paths, while the thick (double-side and colored) arrows denote the directions of sensed 
		subject motions.}
	\label{fig:teaser}
	\vspace{-1.5ex}
\end{figure}
Among all constraints imposed  by Wi-Fi's bistatic nature, we focus on three prominent ones illustrated in Fig.~\!\ref{sfig:teaser_wifi}.\footnote{\textit{Device-based} Wi-Fi sensing~\cite{SpotFi-SIGCOMM15,Chronos-NSDI16,DLoc-MobiCom20} is a special type of bistatic sensing that aims to locate the Tx. Therefore, our ISAC revision to Wi-Fi is orthogonal to this type of sensing applications dedicated solely to localization.} First of all, given the uncertainties such as the existence of carrier frequency offset and the lack of synchronization between Tx and Rx, even estimating the \textit{time-of-flight} (ToF) of the line-of-sight (LoS) path between Tx and Rx entails a very cumbersome process~\cite{Chronos-NSDI16,lin2021spatial}. Unfortunately, the ToFs of the non-LoS (NLoS) paths, albeit essential to device-free sensing~\cite{Widar2-MobiSys18,mDTrack-MobiCom19} 
simply \textbf{cannot} be estimated, hence forcing most of the proposals to be capable of sensing only a single target~\cite{WiAG-MobiSys17,EI-MobiCom18,Widar3-MobiSys19,xiao2021onefi}.
Moreover, whereas 
estimating \textit{angle-of-arrival} (AoA) 
and motion become the centre of Wi-Fi sensing
due to the inability of obtaining ToF, the strong signal of the (useless) LoS path could easily \textbf{overwhelm} the essential NLoS paths. Consequently, existing proposals often have to rely on multiple Wi-Fi links (i.e., multistatic setting)~\cite{Widar-MobiHoc17,WiPose-MobiCom20} and/or on motion effect as an extra hint~\cite{Widar-MobiHoc17,Widar2-MobiSys18}. Finally, it is well known that the motion effect captured by a reflected RF signal represents the distance variation of a reflecting subject along a certain direction. According to Fig.~\!\ref{sfig:teaser_wifi}, this direction happens to be the gradient of the Fresnel 
field~\cite{Fresnel-UbiComp18}, yet this gradient (along which the reflection path length changes) varies with the (unknown) location of the reflecting subject due to the bistatic nature of Wi-Fi, thus causing \textbf{ambiguity} in interpreting the motion sensing results.

Fortunately, all the aforementioned constraints can be lifted if the sensing mode can be converted to \textit{monostatic}: the antenna of each Wi-Fi RF-chain, while transmitting data packets, also captures the reflected signals induced by the transmissions and certain reflecting subjects, as shown in Fig.~\ref{sfig:teaser_isacfi}. Apparently, the ToFs of these reflection paths can be readily obtained as all uncertainties are removed thanks to the co-location of Tx and Rx. Moreover, the AoA and motion of a reflection path can be more accurately estimated without the LoS path interference, exploiting the MIMO (multiple-input and multiple-output) capability of a Wi-Fi device (i.e., its antenna array). Given both ToF and AoA, locating a reflecting subject can be achieved by only one device, yet one may further improve the sensing (for both localization and motion) accuracy by leveraging the interaction between a pair of communicating devices (a distributed MIMO setting).\footnote{It is worth noting that adding the monostatic mode to Wi-Fi sensing operations maximizes the utilization of radio frequency resources, as opposed to the serious wastes under the bistatic-only setting, because two devices in the latter setting obtain far less information than the same two in the former. Essentially, combining both monostatic and bistatic modes can only improve the sensing accuracy from the perspective of estimation theory~\cite{Kay-Est}, thanks to the introduced \textit{diversity gain}.} Last but not least, the interpretation of any motion effect sensed over a reflection path is clearly defined without any ambiguity.

\begin{figure}[t]
	\setlength\abovecaptionskip{8pt}
	\centering
	\subfloat[Conventional Wi-Fi.]{
		\begin{minipage}[b]{0.98\linewidth}
			\centering
			\includegraphics[width = \textwidth]{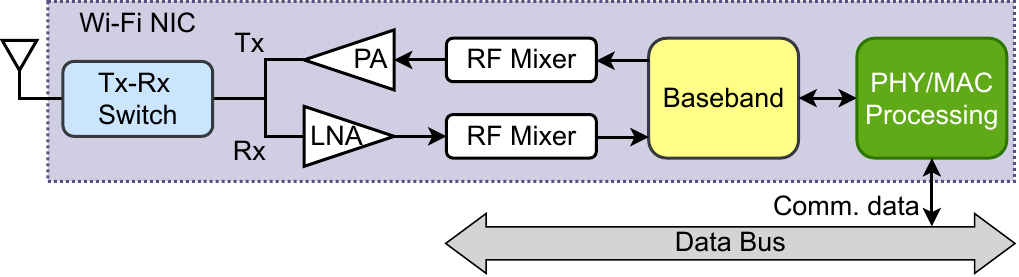}
			\label{sfig:conv_wifi}
			\vspace{-1.5ex}
		\end{minipage}
	}
	\\
	%
	\subfloat[ISAC-Fi full version.]{
		\begin{minipage}[b]{0.98\linewidth}
			\centering
			\includegraphics[width = \textwidth]{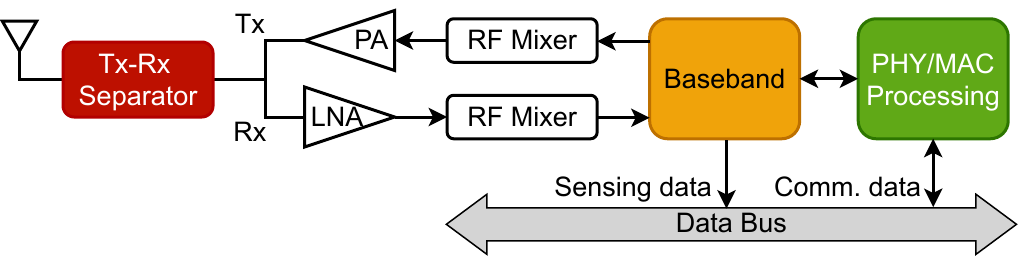}
			\label{sfig:isac_full}
			\vspace{-1.5ex}
		\end{minipage}
	}
	\\
	%
	\subfloat[ISAC-Fi partial version.]{
		\begin{minipage}[b]{0.98\linewidth}
			\centering
			\includegraphics[width = \textwidth]{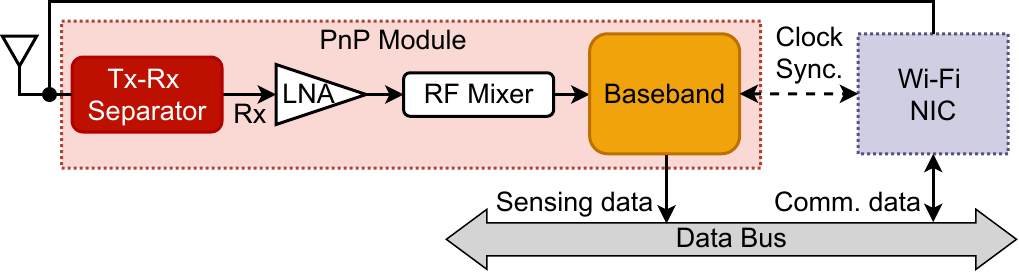}
			\label{subfig:isac_sync_arch}
			\vspace{-1.5ex}
		\end{minipage}
	}
	\vspace{.5ex}
	\caption{Architectures of Wi-Fi (a), full ISAC-Fi (b), and partial ISAC-Fi (c). The hardware novelties mainly lie in replacing the Tx-Rx Switch with a Separator to enable concurrency and in revising the Baseband for enhancing the quality of reflected Rx signals.}
	\label{fig:high_level_archs}
	\vspace{-1.5ex}
\end{figure}

Of course, exploiting monostatic sensing to enable \textit{integrated sensing and communication} (ISAC) within Wi-Fi framework
is far from straightforward; it faces three major challenges. First, the cost of removing LoS path interference is the self-interference from Tx to its own Rx (of the reflected Tx signals) within the same RF chain. Normal radars rely on ultra-wide bandwidth (hence nanosecond time resolution) to separate this Tx-interference~\cite{Octopus-MobiCom21}, which may not be available to Wi-Fi in the current or the next few generations. Second, fully addressing the first problem would entail a revamp of the conventional Wi-Fi hardware configuration (Fig.~\ref{sfig:conv_wifi}), 
upgrading its front-end to handle the Tx-interference (Fig.~\ref{sfig:isac_full}); yet directly implementing this with a commodity Wi-Fi NIC (network interface card) is nearly impossible.
Third, though preserving the Wi-Fi MAC protocol is of primary importance for the sake of compatibility, minor yet critical tuning of the protocol details may be inevitable to, for example, toggle between sensing Rx and communication Rx.
%

To this end, we propose \name as the first trial of enabling ISAC within the Wi-Fi framework. Essentially, we achieve Tx-Rx separation within the same RF chain so as to operate monostatic sensing over Wi-Fi communications; though derived from full-duplex radios~\cite{FDR-SIGCOMM13,WiDeo-NSDI15}, our revision is critical as the original proposals fail to work under ISAC settings. We also work out two prototypes of \name: while \textit{full} \name makes use of USRP X310~\cite{USRP_x310} to emulate a future implementation of Wi-Fi NIC shown in Fig.~\ref{sfig:isac_full}, \textit{partial} \name applies a \textit{plug-and-play} (PnP) module to an arbitrary Wi-Fi NIC, delivering a backward compatibility while rendering conventional NICs ISAC-ready. Finally, we fine-tune the existing Wi-Fi MAC protocols so that both individual and distributed sensing are fully operational without affecting conventional Wi-Fi communications. In summary, we make six major contributions in this paper:
\begin{itemize}
	\item We propose \name as a Wi-Fi based ISAC prototype; it offers, for the first time, the monostatic sensing mode in addition to the commonly adopted bistatic mode.
	%
	\item We design a novel RF front-end to replace the current Wi-Fi design, in order to effectively separate the concurrent sensing and communication signals.
	\item We propose critical revisions to both Wi-Fi MAC protocol and sensing algorithms for maintaining compatibility.
	\item We implement a full prototype of \name leveraging the universal emulation capability of USRP X310.
	\item We also implement a partial prototype of \name; it attaches a PnP module to any existing Wi-Fi NIC in order to elevate it to be ISAC-ready.
	\item We perform extensive experiments with both prototypes to validate their effectiveness and also to demonstrate their superiority over existing Wi-Fi sensing proposals.
\end{itemize}

The rest of our paper proceeds as follow. We first motivate our ideas by exposing the weaknesses of existing Wi-Fi sensing
in Sec.~\ref{sec:problem}. Then we explain our design of the novel RF front-end and the two prototypes of \name in Sec.~\ref{sec:isac-fi}. We further validate the individual functionalities of both prototypes in Sec.~\ref{sec:bench} and compare them with existing Wi-Fi sensing proposals in Sec.~\ref{sec:eval}. We briefly discuss a few related proposals, along with limitations of \name in Sec.~\ref{sec:survey}. Finally, we conclude the paper in Sec.~\ref{sec:conclusion}.

\section{Analysis and Motivations} \label{sec:problem}
In this section, we provide basic theoretical and experiment analyses to compare bistatic Wi-Fi sensing with the novel monostatic mode in terms of channel model; these analyses and comparisons serve as motivations and inspirations for the design of our \name. 
Basically, the Wi-Fi OFDM signal $x(t, \tau)$ received over the air and modulated onto a certain carrier frequency $f_{\mathrm{c}}$ is given by:
\begin{align}  \label{eq:rf_csi}
	x(t, \tau) = \underbrace{ \sum_{p = 0}^{M} \alpha_{p} \delta \left(t - \tau_p - \tau_p^{\mathrm{D}}(t)\right)}_{\bm{h(t, \tau)}:~\text{channel over the air}} \ast \underbrace{ e^{-j2 \pi f_{\mathrm{c}} t}  s(t)}_{\substack{\text{Tx \textbf{baseband}} \\ \text{symbol}~s(t)~\text{after} \\ \text{up-conversion}}},
\end{align}
where the symbol $\ast$ refers to convolution; $\tau_p$ and $\tau_p^{\mathrm{D}}(t)$ denote the propagation delay and the motion-induced delay along the $p$-th propagation path, respectively: they are the \textit{key sensing information} offered by Wi-Fi communications.

\subsection{Uncertainties in Temporal Features} \label{ssec:tfchannel}
We characterize the uncertainties of the temporal features in a channel model, and then discuss their implications. 

%
\paragraph{Modelling Offsets} 
%
Since the crystal oscillators (i.e., clocks) of Tx and Rx may differ slightly, the resulting imperfect signal processing introduces several random offsets to contaminate both $\tau_p$ and $\tau_p^{\mathrm{D}}(t)$. 
To understand the details of these errors, let us walk through the whole processing line of Rx chain. 
First of all, down-converting the OFDM signal $x(t, \tau)$ in the Rx chain requires applying $e^{j2 \pi f_{\mathrm{c}}t }$ to shift $x(t, \tau)$ to baseband, but the resulting baseband signal is actually:
\begin{align}  \label{eq:bb_csi_fre_err}
	y(t, \tau) = {  h(t,\tau )}  \ast { e^{-j2 \pi  \left( \gamma_{\mathrm{c}} t + \phi_{\mathrm{c}} \right)  }  s(t)},
\end{align}
where $\gamma_{\mathrm{c}} =  f_{\mathrm{c}} - f_{\mathrm{c}}'$ denotes the CFO (Carrier Frequency Offset) caused by the residue error in PLL (Phase Locked Loop); it forces Rx to match $f_{\mathrm{c}}$ with a slightly different $f_{\mathrm{c}}'$. Moreover, a CPO (Carrier Phase Offset) $\phi_{\mathrm{c}}$ is imposed by both PLL and VCO (Voltage Controlled Oscillator), since VCO has a random phase each time it starts or restarts and PLL cannot fully compensate the phase difference between the local Rx carrier and the received signals $x(t,\tau)$. 


Further down the processing line, the baseband signal $y(t, \tau)$ is sampled by ADC and then converted to frequency domain via FFT. Considering an OFDM symbol with size $N_{\mathrm{FT}}$ and Rx sampling period $T_{\mathrm{s}} = f_{\mathrm{s}}^{-1}$ with $f_{\mathrm{s}}$ being the sampling rate, we let $t=n T_{\mathrm{s}}$ with $n$ denoting the sampling index and thus obtain the following $k$-th sub-carrier signal of the $\ell$-th OFDM symbol after FFT~\cite{OFDM_TCOM94}:
\begin{align}  \label{eq:bb_csi_time_err}
	Y_\ell( k , \tau) = {  H^\star S( k )}  { e^{-j2 \pi  \left (\ell \frac{ \gamma_{\mathrm{c}}}{ \Delta f N_{\mathrm{FT} }  }    + \phi_{\mathrm{c}} \right )  }  e^{-j 2 \pi  \frac{k \beta}{N_{\mathrm{FT}} }     }    },
\end{align}
%
%
where $H^\star = H(k, \tau ) = \sum_{p=0}^{M} \alpha_{k,p} e^{-j 2 \pi f_{\mathrm{c}} (\tau_p + \tau_p^{\mathrm{D}}) }$,  $\Delta f$ is the OFDM sub-carrier spacing, and $\beta = \frac{T_{\mathrm{s}} - T_{\mathrm{s}}' }{ T_{\mathrm{s}}' }$, with $T_{\mathrm{s}}'$ denoting the Tx sampling period, is the SFO (Sampling Frequency Offset) caused by the difference between Tx DAC and Rx ADC clocks. 

Finally, since the lack of knowledge on the starting point of an OFDM symbol at the Rx side, it is hard to determine the right samples to feed into FFT. This issue persists even with a carefully designed preamble and corresponding detection algorithms~\cite{OFDM_TCOM94},
causing a phase error because missing even a small length of the preamble equivalently results in a non-negligible delay. We term this phase error PDD (Packet Detection Delay) $\epsilon$; it necessitates a revision to $Y_\ell(k, \tau)$:
\begin{align}  \label{eq:bb_csi_time_err_pdd}
	Y_\ell( k , \tau) =
	{ H^\star S( k )}  { e^{-j2 \pi  \left (\ell \frac{ \gamma_{\mathrm{c}}}{ \Delta f N_{\mathrm{FT} }  }    + \phi_{\mathrm{c}} \right )  }  e^{-j 2 \pi  \frac{k \beta}{N_{\mathrm{FT}} }}   
	e^{-j 2 \pi   \frac{k \epsilon}{N_{\mathrm{FT}}}}}. 
\end{align}
Though all these errors exist in normal Wi-Fi communications, they have been masked by well-designed demodulation schemes. However, sensing aims to capture minor variations, rendering it intolerable to even minor errors and hence fundamentally different from communication.

\begin{figure}[b]
	\vspace{-2ex}
	\raggedleft
	\!\!\subfloat[Bistatic mode.]{ 
		\begin{minipage}[b]{.49\linewidth}
			\centering
			\includegraphics[width = \textwidth]{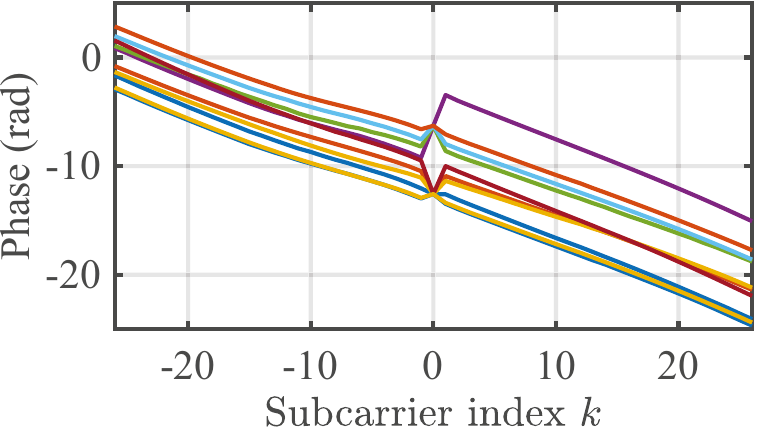}
			\label{sfig:csi_phases}
			\vspace{-1.5ex}
		\end{minipage}
	}
	\subfloat[Monostatic mode.]{ 
		\begin{minipage}[b]{.49\linewidth}
			\centering
			\includegraphics[width = \textwidth]{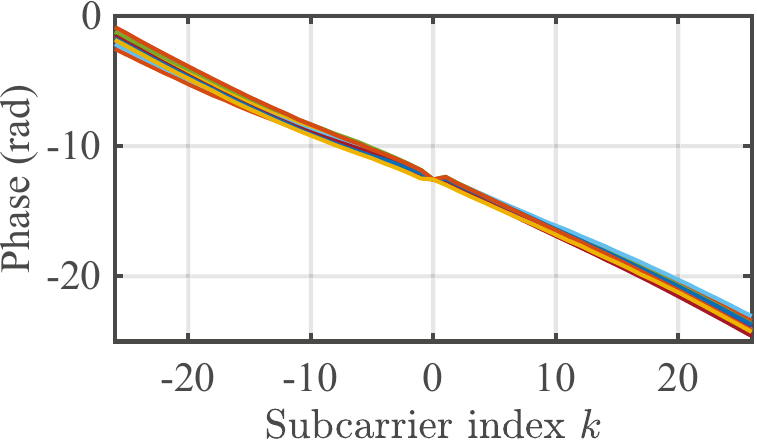}
			\label{sfig:diff_csi_phases}
			\vspace{-1.5ex}
		\end{minipage}
	}
	\caption{The unwrapped CSI phases of 52 subcarriers and across consecutive symbols (marked with different colors) under bistatic (a) and monostatic (b) modes.}
	\label{fig:hardware_offsets}
\end{figure}

\paragraph{Bistatic vs. Monostatic Sensing.} Apparently, all uncertainties in the channel model Eqn.~\eqref{eq:bb_csi_time_err_pdd} (i.e., CFO, CPO, SFO, and PDD) affect bistatic sensing. Therefore, it is extremely challenging (if it is ever possible) to measure quantities induced by temporal features (e.g., ToF from $\tau_p$). 
%
On the contrary, switching to the monostatic mode so that Tx and Rx become co-located in the same device, they would share the same clock. Therefore, CFO, SFO, and PDD can be significantly reduced. Though CPO still persists, obtaining it during the hardware initialization is viable.
%
%
We measure the CSI phases of the same symbol in consecutive packets under both bistatic and monostatic modes in an empty room. As shown in Fig.~\!\!\ref{fig:hardware_offsets}, the phases under the bistatic mode increase gradually with $\ell$ (symbol) and have
smaller slopes in $k$ (subcarrier) than those under the monostatic mode, which accords well with the phase terms in Eqn.~\eqref{eq:bb_csi_time_err_pdd} given negative $\beta$ and $\epsilon$. 
The monostatic mode, on the contrary, exhibits only minor phase variations across both $\ell$ and consistent slops in $k$, thus allowing for accurate recovery of temporal features $\tau_p$ and $\tau_p^{\mathrm{D}}(t)$. The glitches at the 0-th subcarrier in Fig.~\ref{fig:hardware_offsets} are caused by lack of data stream
and the phase unwrapping process, which may jump drastically due to CFO under the bistatic mode but are well controlled otherwise.
%

\subsection{Dominating Interference from LoS Path} \label{sec:ssec:dpi}
According to the channel model in Eqn.~\eqref{eq:rf_csi}, multiple signal propagation paths exist, among them two are special: except the 0-th path to be clarified later, the 1-st path (the LoS path between Tx and Rx) has a dominating power over all other NLoS paths under the bistatic mode, yet it disappears under the monostatic mode. To better understand the LoS path interference to NLoS paths, we expand the amplitude $\alpha_{p}$ in Eqn.~\eqref{eq:rf_csi} to calculate the received power $P_{p}^{\mathrm{Rx}}$ at an Rx antenna~\cite{BPath-RADAR03}:
\begin{align}  \label{eq:rxPower}
	P_{p}^{\mathrm{Rx}} =\alpha_{p}^{2} = \frac{P^{\mathrm{Tx}} G^{\mathrm{Tx}} G^{\mathrm{Rx}  } \lambda_{\mathrm{c}}^2 \sigma_p  }{ (4 \pi)^3 (R_p^{\mathrm{Tx}} R_p^{\mathrm{Rx}} )^2} ,
\end{align}
where $P^{\mathrm{Tx}}$ is the Tx power, $G^{\mathrm{Tx}}$ and $G^{\mathrm{Rx}  }$ are the Tx and Rx antenna gains, $\lambda_{\mathrm{c}}$ denotes the wavelength of the carrier frequency,  $ \sigma_p$ represents the RCS (Radar Cross Section) of the reflecting target, and $R_p^{\mathrm{Tx}}$ and $R_p^{\mathrm{Rx}}$ represent the Tx-target and target-Rx ranges. 
As far as the target does not lie on the LoS path (which is very rare), the power ratio between the $p$-th (NLoS) path signal carrying the (reflected) sensing information and the interfering LoS path signal becomes:
\begin{align}  \label{eq:rxPowerRatio}
	\eta_p =  \frac{P_{p}^{\mathrm{Rx}} }{P_{1}^{\mathrm{Rx}} } = \frac{L \sigma_p }{ 4 \pi (R_p^{\mathrm{Tx}} R_p^{\mathrm{Rx}} )^2},
\end{align}
{where $L$ denotes the LoS distance between Tx and Rx.}

Assuming a bistatic sensing with $L = 2$~\!m,  $\sigma_p = 1\!~\text{m}^2$, and $R_p^{\mathrm{Tx}} = R_p^{\mathrm{Rx}} = 2$~\!m just for simplicity, Eqn.~\eqref{eq:rxPowerRatio} suggests that $\eta_p \approx -40\!~\mathrm{dB} $. 
We should be reminded that, as the LoS path signal is meant for communications (the main function of Wi-Fi), there is no way to suppress it just for sensing purpose.
Fortunately, operating the sensing function under the monostatic mode could totally remove the LoS path; in other words, the constraint imposed by Eqn.~\eqref{eq:rxPowerRatio} disappears. In fact, under the monostatic mode, the Tx-target-Rx round-trip path becomes the dominating one (see Fig.~\!\ref{sfig:teaser_isacfi}), and it happens to carry the desired sensing information. 
\begin{figure}[b]
	\setlength\abovecaptionskip{8pt}
	\vspace{-3.5ex}
	\raggedleft
	\!\!\subfloat[{Bistatic mode.}]{ 
		\begin{minipage}[b]{.49\linewidth}
			\centering
			\includegraphics[width = \textwidth]{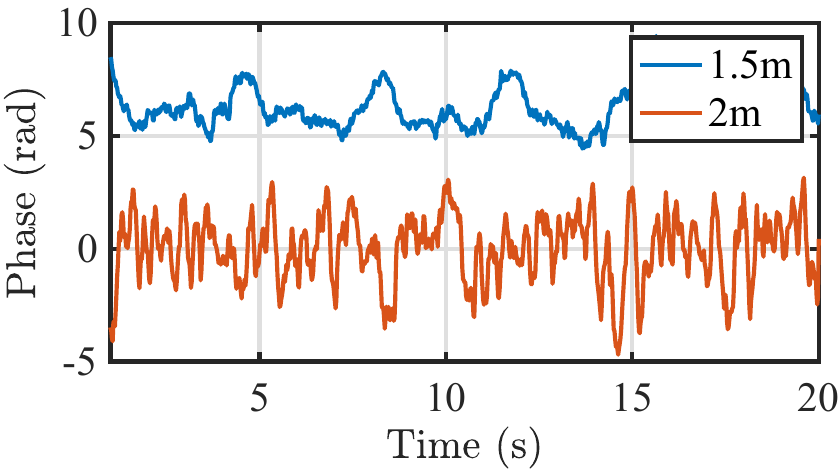}
			\label{sfig:bi_los}
			\vspace{-2.5ex}
		\end{minipage}
	}
	\subfloat[Monostatic mode.]{ 
		\begin{minipage}[b]{.49\linewidth}
			\centering
			\includegraphics[width = \textwidth]{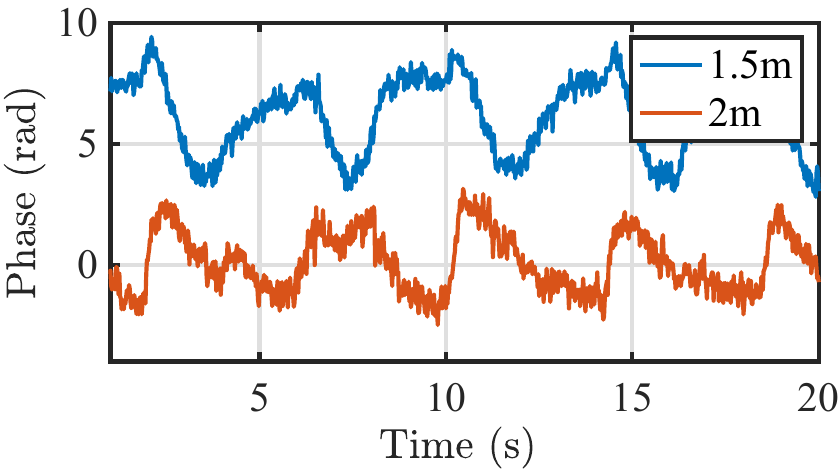}
			\label{sfig:mono_los}
			\vspace{-2.5ex}
		\end{minipage}
	}
	\caption{The phase variations induced by human (target) breath at two different Rx-target ranges under bistatic (a) and monostatic (b) modes.}
	\label{fig:los_dominating}
\end{figure}
We use a set of experiments to briefly demonstrate the differences, where we set $L=1.5$~\!m and vary the target-Rx ranges. As shown in Fig.~\ref{fig:los_dominating}, the LoS interference is very evident under the bistatic mode, especially when compared with those under the monostatic mode. Nonetheless, we do face a new challenge under the monostatic mode: the $0$-th path (or \textit{Tx-interference}) signal in Eqn.~\eqref{eq:rf_csi}, absent under the bistatic mode due to the temporal separation enforced by CSMA/CA MAC protocol, will cause a serious problem. This new challenge is certainly the key issue to be tackled in our paper.

\subsection{Ambiguity in Motion Sensing} \label{ssec:ambiguity-motion}

Let us now focus on the motion-induced delay $\tau_p^{\mathrm{D}}$ in Eqn.~\eqref{eq:rf_csi}: it is a quantity representing the variations (e.g., target motion) along the $p$-th path. Basically, $\tau_p^{\mathrm{D}}(t) = \frac{ \Delta R_p(t) }{\mathbf{c}} $ where $\mathbf{c}$ is the speed of light and $\Delta R_p(t)$ is the instantaneous variation in range at time $t$. Though $\Delta R_p(t)$ is often termed \textit{displacement} in radar terminology, it is actually a scalar obtained by projecting the actual displacement of a moving target onto a certain \textit{direction}. Whereas this direction can be readily characterized
under the monostatic mode (the radial direction from the Tx/Rx shown in Fig.~\!\ref{sfig:teaser_isacfi}), it is nontrivial to determine and hence ambiguous under the bistatic mode.

\begin{figure}[t]
	\centering
	\includegraphics[width=.7\linewidth]{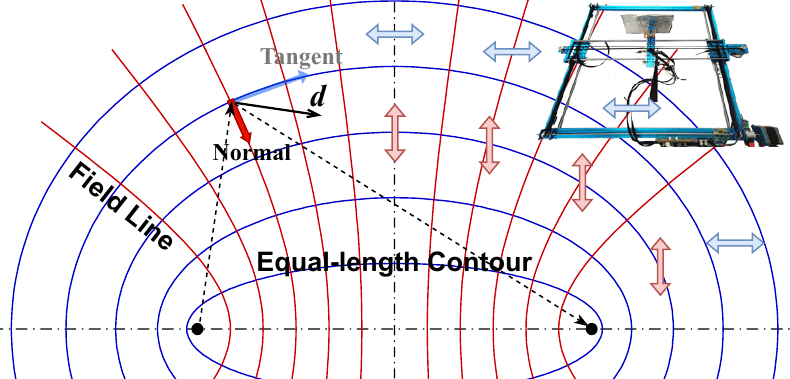}
	\caption{Sensing motion effect under bistatic mode: conceptual illustration and experiment settings.}
	\label{fig:doppler_exp}
	\vspace{-2ex}
\end{figure}

Because $\Delta R_p(t)$ represents the variation in range and the range is actually the length of Tx-target-Rx reflection path under the bistatic mode, we can define a \textit{field} with Tx and Rx as two \textit{focus point}s. As this specific field describes the lengths of Tx-target-Rx reflection paths, its \textit{equipotential surface}s correspond to equal-length contours that happen to be ellipsoids with Tx and Rx as foci (see Fig.~\!\ref{fig:doppler_exp}). At any point in the field, a target displacement $\vec{d}$ can be decomposed into tangent and normal components based on the ellipsoid on which the target resides. Since $\Delta R_p(t)$ senses the variant in range, it can only represent the normal component whose direction varies with the target location, whereas the tangent component leads to variation along an equal-length contour and thus delivers no impact on $\Delta R_p(t)$. Further reasoning could deduce that all normal directions lie on hyperbolas confocal with (hence orthogonal to) the ellipsoids, which can be deemed as the \textit{field line}s of this field.

\begin{figure}[t]
\setlength\abovecaptionskip{8pt}
\vspace{-2ex}
\raggedleft
\!\!\subfloat[Perpendicular.]{ 
	\begin{minipage}[b]{.49\linewidth}
		\centering
		\includegraphics[width = \textwidth]{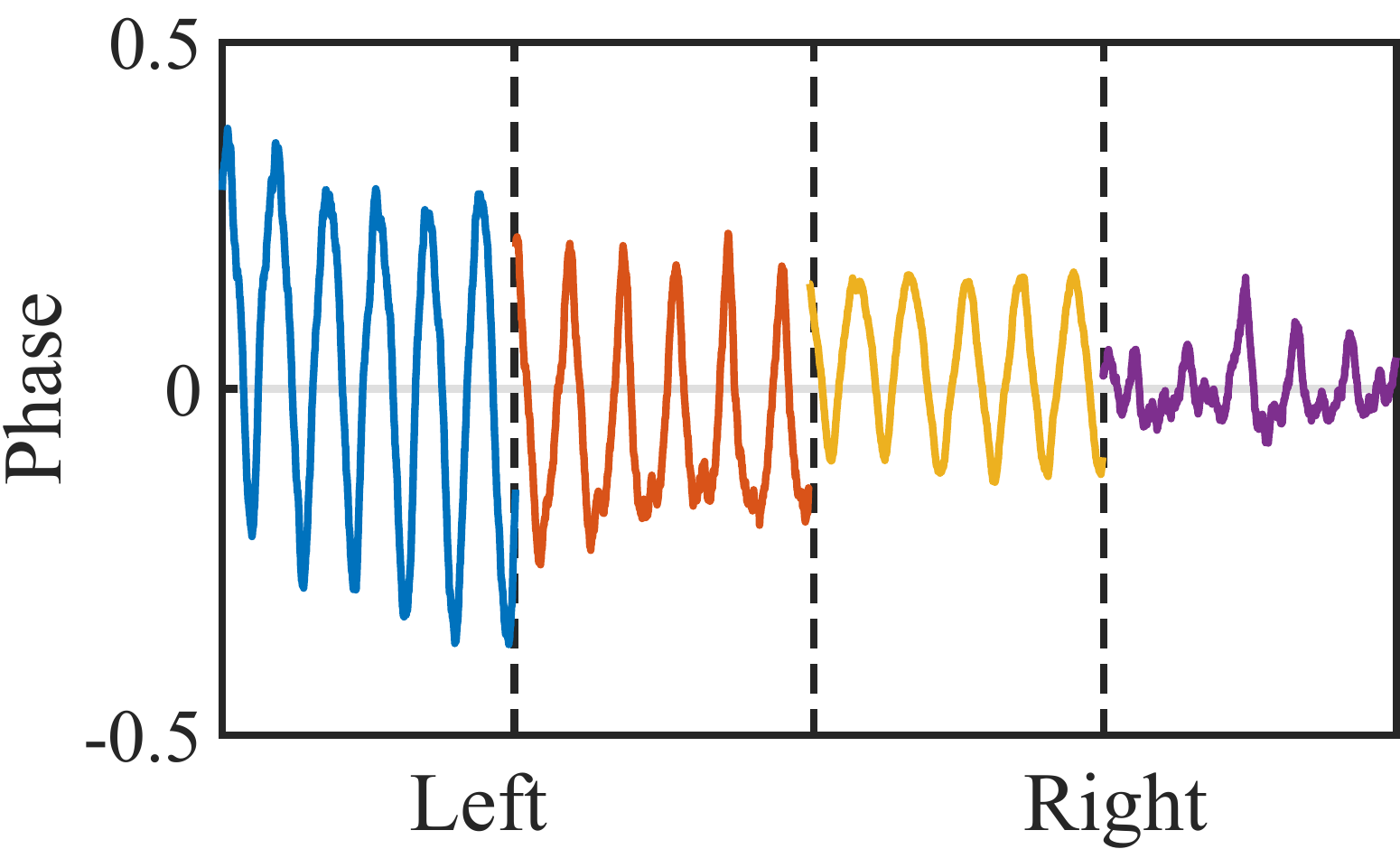}
		\label{sfig:perpendicular}
		\vspace{-1.5ex}
	\end{minipage}
}
\subfloat[Parallel.]{ 
	\begin{minipage}[b]{.49\linewidth}
		\centering
		\includegraphics[width = \textwidth]{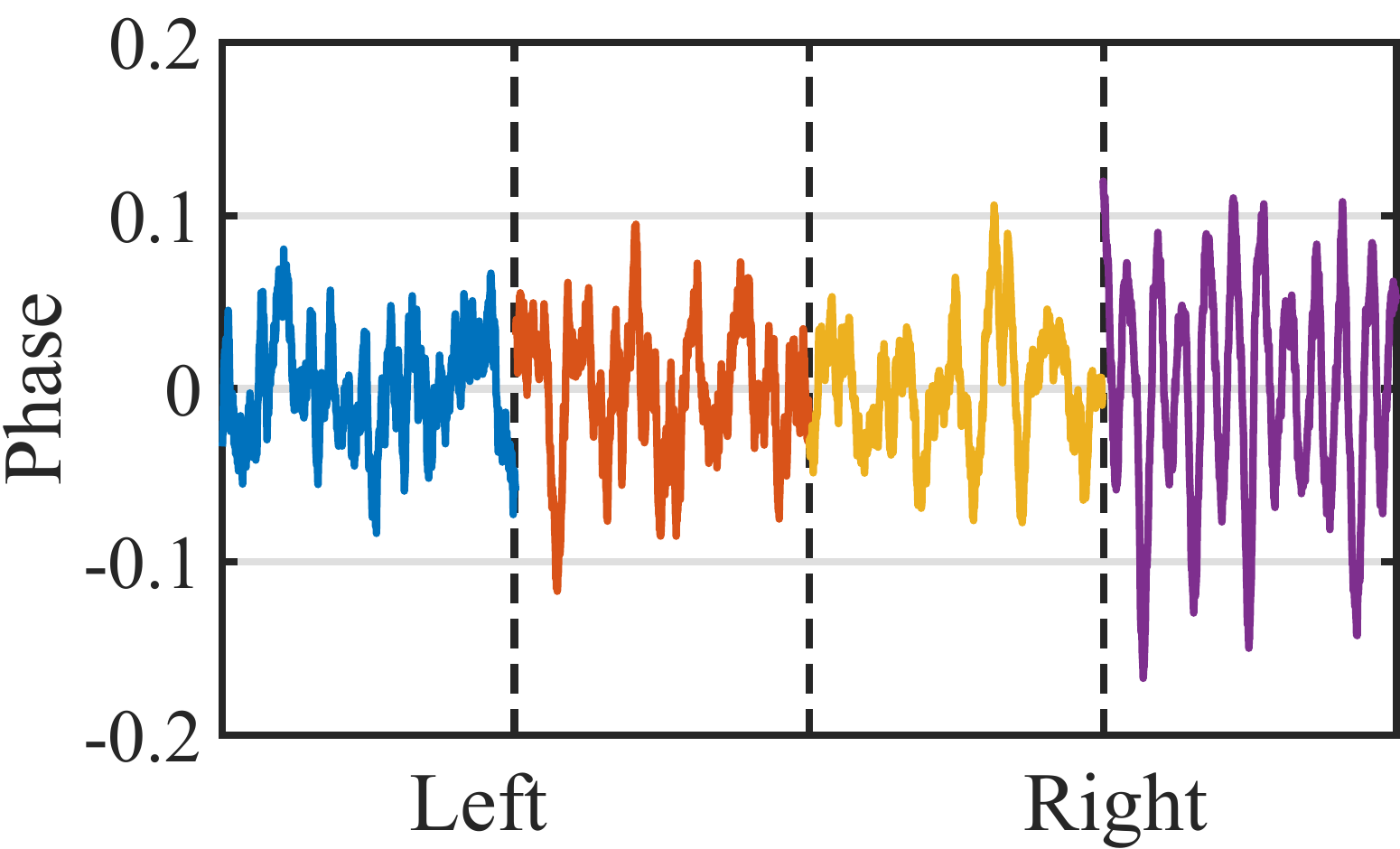}
		\label{sfig:parallel}
		\vspace{-1.5ex}
	\end{minipage}
}
\caption{The unwrapped CSI phases of the 1-st subcarrier when a motor-driving slide rail is placed perpendicular (a) and parallel (b) to the Tx-Rx line. Different positions of the rail are marked by distinct colors.}
\label{fig:ambiguity_motion}
\vspace{-1.5ex}
\end{figure}

Although it is highly nontrivial to experimentally characterize this field, verifying the ambiguity in sensing motion direction can be readily obtained by well-controlled experiments. Specifically, we adopt a motor-driving slide rail programmable to move a target in a constant speed. According to the aforementioned analysis, putting the rail parallel or perpendicular to the Tx-Rx line (as shown by the thick double-side arrows in Fig.~\ref{fig:doppler_exp}) and varying its position within the field, the sensed $\Delta R_p(t)$ should exhibit magnitude variations even though the target is programmed to have a constant speed along the rail, simply due to the monotonically varying projections onto the field lines. As shown in Fig.~\ref{fig:ambiguity_motion}, the monotonic trends of $\Delta {R_p}(t)$ (represented by phases) are evident under both perpendicular and parallel cases, firmly corroborating our earlier analysis.
%
%

\paragraph*{Remarks} It is worth noting that enabling the monostatic mode in Wi-Fi does not replace its originally bistatic sensing ability, because communications are still half-duplex as defined by CSMA/CA MAC. Instead, it simply exerts the full potential of Wi-Fi sensing over communications: instead of having a pair of Wi-Fi devices working as one bistatic radar, we can simultaneously have two monostatic and one bistatic radars. Nonetheless, this ISAC architecture on Wi-Fi entails the need for a fine-tuning of the MAC protocol to differentiate the Rx status under different radar modes.
%
%
%

\section{ISAC-F{\MakeLowercase{i}}: Making Wi-Fi ISAC-Ready} \label{sec:isac-fi}
We explain the design of \name in four steps, started by a brief overview. We first explain how to realize the Tx-Rx separator as the basic enabler of \name (for both full and partial prototypes), then 
we discuss the potential issues and corresponding countermeasures for both co-existence with Wi-Fi framework and channel parameter estimation under irregular traffic. Finally, we elaborate on the implementation of collaborative MIMO sensing under \name.

\subsection{System Overview}
\begin{figure}[b]
\setlength\abovecaptionskip{8pt}
\vspace{-.5ex}
\centering
\includegraphics[width=.96\linewidth]{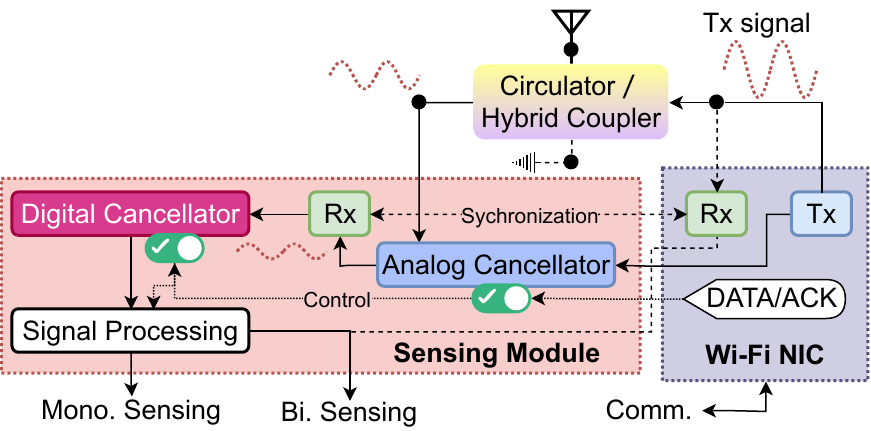}
\caption{Tx-Rx separation for \name.}
\label{fig:high_level_arch}
\vspace{-1ex}
\end{figure}

%
%
The hardware design of \name is centered around the ability of separating concurrent Tx and Rx signals. Therefore, we use Fig.~\ref{fig:high_level_arch} to capture the essential implementation details of the Tx-Rx separator, and we briefly explain this seemingly complicated structure with several key points. 

\begin{itemize}
\item This structure represents two complementary designs. A \textit{full} version integrates both sensing and communication functions, so two {\small\textsf{Rx}} chains are merged into one and all signal paths represented by dashed lines disappear. However, as the full version requires revamping the design of Wi-Fi NICs, we also offer a \textit{partial} version with backward compatibility: it appends a special {\small\textsf{Sensing Module}} to an arbitrary {\small\textsf{Wi-Fi NIC}}.
\item Though inspired by full-duplex radios~\cite{FDR-SIGCOMM13}, our Tx-Rx separator differs significantly from it in that the Rx (monostatic sensing) signal to be extracted is, instead of arbitrary Rx signal, the reflection of the (slightly earlier) Tx signal. For the full version, we adopt a {\small\textsf{Circulator}} at the first stage to physically separate Tx and Rx, whereas we employ a {\small\textsf{Hybrid Coupler}} in the partial version to isolate Rx signal before feeding it to the sensing module. Other components for suppressing the residue Tx-interference are shared by both versions. We elaborate the two cancellators in Sec.~\ref{ssec:separator}, with an emphasis on preserving the monostatic sensing signal.
\item The sensing module has to be compatible with the Wi-Fi framework; in particular, the {\small\textsf{Analog/Digital Cancellator}}s should not operate under the reception (hence bistatic) mode. To this end, we leverage the {\small\textsf{DATA/ACK}} messages to invoke transitions among three major states of \name: {\small\textsf{Communication}}, {\small\textsf{Monostatic Sensing}}, and {\small\textsf{Bistatic Sensing}}, as presented in Sec.~\ref{ssec:protocol}.
%
%
\item After the Tx-Rx separation, a special {\small\textsf{Signal Processing}} procedure is applied to treat bistatic and monostatic signals separately. Since existing Wi-Fi sensing proposals 
assume regular data packets that are far from realistic,
we discuss how to make sensing compatible with irregular data packets in Sec.~\ref{ssec:monochannel}.
%
%
%
\end{itemize}
With the above points concerning only single-device operations of \name, we stress that, given a proper information sharing scheme provided by both the Wi-Fi and backbone (wired) networks, a set of Wi-Fi devices (APs and NICs) can collaboratively serve as a distributed MIMO sensing system. We briefly discuss a possible protocol design to facilitate this collaborative sensing paradigm in Sec.~\ref{ssec:collaborate}.
%

\subsection{Tx-Rx Separator Design} \label{ssec:separator}
As the circulator and hybrid coupler are both commodity components, we only explain the details of the two cancellators. Because the monostatic sensing signals are reflected version of the Tx signals, they would be treated as interference from the perspective of full-duplex radios~\cite{FDR-SIGCOMM13}. Consequently, our main contribution lies in preserving these sensing signals while removing the Tx-interference. In particular, we start with two preliminary designs of the cancellators, then their inherent problems are analysed and hence revised to achieve our self-adapted Tx-Rx separation.

\paragraph{Analog Cancellator} 

\begin{figure}[t]
\setlength\abovecaptionskip{8pt}
\centering
\includegraphics[width=\linewidth]{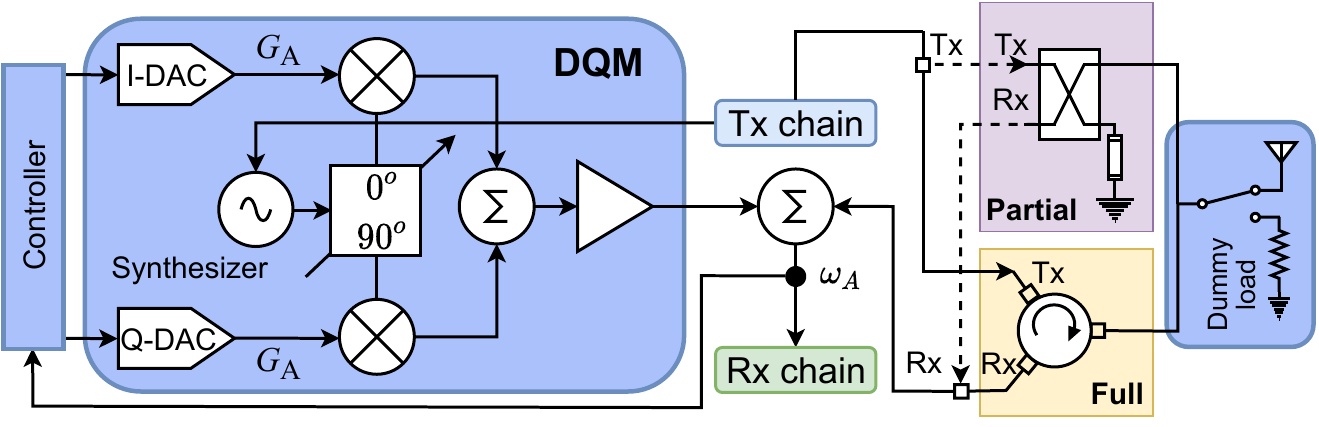}
\caption{The details of the analog cancellator.}
\label{fig:analog_cancellator}
\vspace{-1ex}
\end{figure}

This component takes the output of the circulator or hybrid coupler as its analog input. 
Since the received signal contains Tx-interference ($p=0$) and multipath reflections ($p>0$) via different channels respectively, the output of the cancellator becomes:
%
%
\begin{align}  \label{eq:analog_sep}
\!\!\!\!z(t, \tau) =  \left[ G_{\mathrm{A}} \cdot x(t, \tau_0) +  G_{\mathrm{H}} \cdot x(t, \tau_0) \right] +  G_{\mathrm{C}} \cdot x(t,\tau_{p > 0}),\!\!
\end{align}
where $G_{\mathrm{A}}$, $G_{\mathrm{H}}$, and $G_{\mathrm{C}}$ denote the channel gains of the analog cancellator, the RF hardware, and the circulator/hybrid coupler, respectively. 
We slightly abuse the terminology by $\tau_p$ denoting both $\tau_p$ and $\tau_p^{\mathrm{D}}$ in Eqn.~\eqref{eq:rf_csi}. Let the residue Tx-interference be $\omega_{\mathrm{A}}(t) = G_{\mathrm{A}} \cdot x(t, \tau_0) + G_{\mathrm{H}} \cdot x(t, \tau_0)$,
the analog cancellator should adjust $G_{\mathrm{A}}$ so as to minimize $\omega_{\mathrm{A}}(t)$. 
Different from the implementation in \cite{FDR-SIGCOMM13}, we adopt Direct Quadrature Modulator~(DQM) to realize the analog cancellator shown in Fig.~\ref{fig:analog_cancellator}. This much simpler yet more effective architecture treats $G_{\mathrm{A}}$ as the inverse of $G_{\mathrm{H}}$ (ideally only comprised of 
antenna and RF circuits), and controls the IQ baseband generator of DQM to match $G_{\mathrm{A}}$ with  $G_{\mathrm{H}}$ in order to minimize $\omega_{\mathrm{A}}(t)$. 
This is more compact and effective (with wider dynamic range and higher resolutions in both amplitude and phase) than the fixed delay circuit in~\cite{FDR-SIGCOMM13}.

\paragraph{Digital Cancellator} 
%
As \name demands only CSI extraction for sensing but has no interest on data contents, we propose a preamble-based digital cancellator: it samples the output preambles from the analog cancellator via correlations, which have passed through the RF downconversion and ADC sampling, thus containing $\omega'_{\mathrm{A}}(nT_{\mathrm{s}})$ as the residue analog Tx-interference: 
%
%
\begin{align}  \label{eq:digital_sep}
z(n T_{\mathrm{s}}, \tau) = \left[ G_{\mathrm{D}} \cdot s(n T_{\mathrm{s}}) + \omega'_{\mathrm{A}}(n T_{\mathrm{s}}) \right] + x(n T_{\mathrm{s}}, \tau_{p > 0}),
\end{align}
where $G_{\mathrm{D}}$ denotes an adaptive filter whose coefficients are obtained via the Least Mean Squares~(LMS) algorithm with low complexity and fast convergence. Consequently,
a linear combination of multiple time-delayed versions of $s(n T_{\mathrm{s}})$ is constructed by applying $G_{\mathrm{D}}$, aiming to offset $\omega'_{\mathrm{A}}(t)$. Here $s(n T_{\mathrm{s}})$ represents the baseband of the known Wi-Fi preamble pre-stored by \name; using it avoids the much larger errors in LMS processing inherent to the data part.

\begin{figure}[t]
\setlength\abovecaptionskip{8pt}
\centering
\includegraphics[width=.68\linewidth]{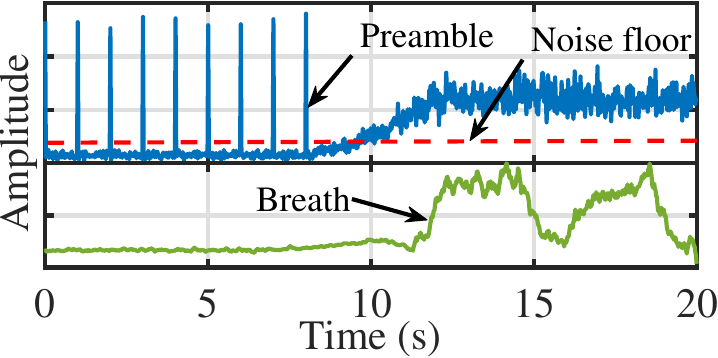}
\vspace{-.5ex}
\caption{Preliminary cancellators erase monostatic sensing signal. The breath signals have been artificially amplified to be more conspicuous.}
\label{fig:tx_rx_sep_ref}
\vspace{-1.5ex}
\end{figure}

\paragraph{Self-Adapted Tx-Rx Separation} 
Though the above two cancellators appear to be plausible, the adjustments to $G_{\mathrm{A}}$ and $G_{\mathrm{D}}$ cannot perfectly focus on the Tx-interference. In practice, as the minimization can only be applied to either $z(t, \tau)$ or $z(n T_{\mathrm{s}}, \tau)$, the cancellators could potentially offset the third term in both Eqn.~\eqref{eq:analog_sep} and~\eqref{eq:digital_sep}. However, the CSIs contained in these terms are valuable information demanded by monostatic sensing. Therefore, the biggest challenge is how to keep $x(n T_{\mathrm{s}}, \tau_{p > 0})$ while removing the Tx-interference. 
We illustrate this challenge 
using an experiment sitting a human subject close to \name (with preliminary cancellators). According to Fig.~\ref{fig:tx_rx_sep_ref}, 
while applying the cancellators (before 8~\!s with the spikes indicating preamble receptions and hence cancellator recalibrations) suppresses the Tx-interference below the noise floor, the human breath signal captured by $x(n T_{\mathrm{s}}, \tau_{p > 0})$ also disappears. On the contrary, stopping the cancellation brings back both the Tx-interference and breath signal, albeit the latter being heavily distorted by the former.
%

\begin{figure}[h]
\setlength\abovecaptionskip{6pt}
\vspace{-2.5ex}
\raggedleft
\!\!\subfloat[Antenna impact is minimum.]{ 
	\begin{minipage}[b]{.49\linewidth}
		\centering
		\includegraphics[width = \textwidth]{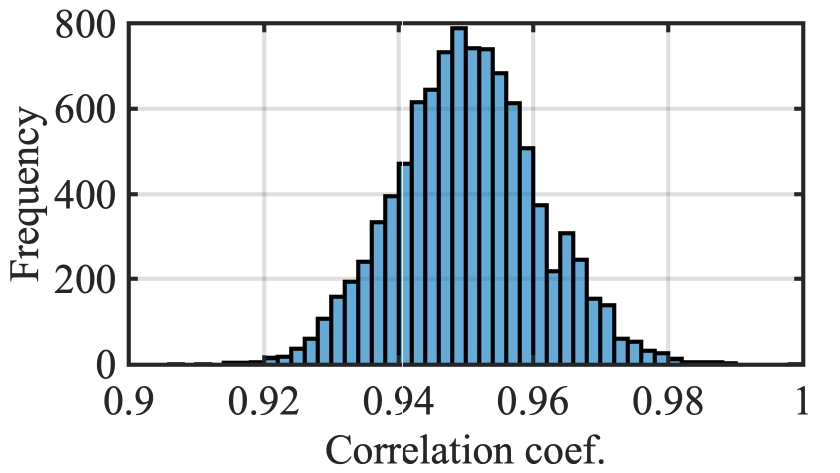}
		\label{sfig:antenna_dummy}
		\vspace{-2.5ex}
	\end{minipage}
}
\subfloat[Signal stable after calibration.]{ 
	\begin{minipage}[b]{.49\linewidth}
		\centering
		\includegraphics[width = \textwidth]{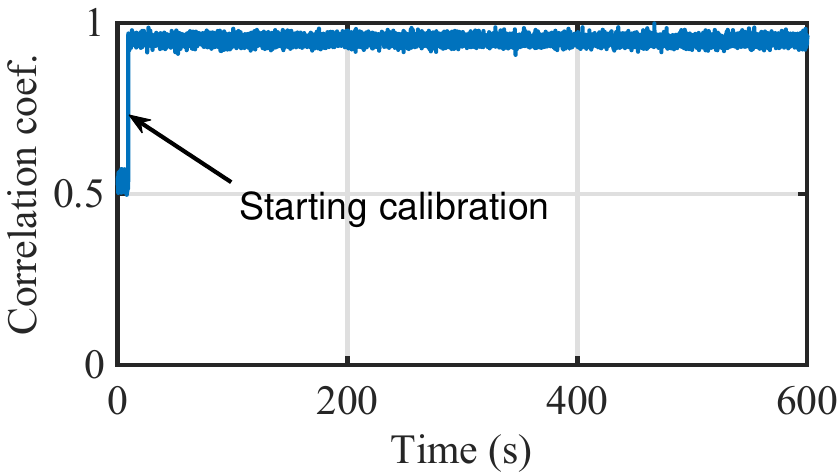}
		\label{sfig:channel_stable}
		\vspace{-2.5ex}
	\end{minipage}
}
\caption{The correlation coefficients of (a) Tx-interferences under two antenna settings and (b) Rx signal right after a calibration and those received later.}
\label{fig:obervations}
\end{figure}

Fortunately, we make two observations potentially resulting in a solution. We first observe that almost the entire Tx-interference $x(t, \tau_0)$ comes from hardware circuits rather than the antenna. We perform two experiments where RF absorbing materials are used to surround an antenna in one case, and the antenna is replaced by an RF dummy load to stop radiating RF signals in another case. We compare the Tx-interference under these two cases in Fig.~\ref{sfig:antenna_dummy}: the correlation coefficients between them are over 90\%, proving that the antenna has almost no impact on the Tx-interference. Our second observation is that the hardware channel of Tx-interference is stable in a long term. Therefore, with proper calibrations on both $G_{\mathrm{A}}$ and $G_{\mathrm{D}}$, they can keep cancelling the Tx-interference without further fine-tuning in at least ten minutes (practical calibration intervals can be made shorter), as shown in Fig.~\!\ref{sfig:channel_stable}. 

\begin{figure}[t]
\setlength\abovecaptionskip{6pt}
\centering
\includegraphics[width=.68\linewidth]{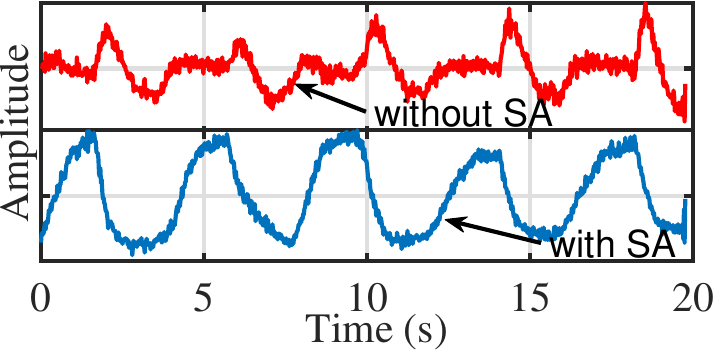}
\vspace{-.5ex}
\caption{The human breath signals with and without SA; SA represents the \textit{Self-Adapted} Tx-Rx separation. }
\label{fig:self-adapted}
\vspace{-2.5ex}
\end{figure}

According to these observations, we decide to add an RF switch to toggle between the antenna and a dummy load in the \name design, so as to realize a self-adapted Tx-Rx separation (the right-most component in Fig.~\ref{fig:analog_cancellator}); this causes only a minor variation bearing negligible complexity and monetary costs. Basically, \name switches its Tx port from the antenna to the dummy load in a regular basis or before enabling monostatic sensing, allowing for properly calibrating both $G_{\mathrm{A}}$ and $G_{\mathrm{D}}$. During monostatic sensing piggybacking on communications, \name switches its Tx port to the antenna and leverages both cancellators ($G_{\mathrm{A}}$ and $G_{\mathrm{D}}$) to suppress the Tx-interference; the concerned state transitions shall be elaborated soon. As shown in Fig.~\ref{fig:self-adapted}, 
the human breath is clearly extracted with the self-adapted Tx-Rx separation; otherwise it can be damaged by residual Tx-interference resulting from incomplete cancellation of normally calibrated cancellators.

\subsection{Co-existing with Wi-Fi Framework} \label{ssec:protocol}
\begin{figure}[h]
\setlength\abovecaptionskip{8pt}
\vspace{-2.5ex}
\raggedleft
\!\!\subfloat[SNR.]{ 
	\begin{minipage}[b]{.49\linewidth}
		\centering
		\includegraphics[width = \textwidth]{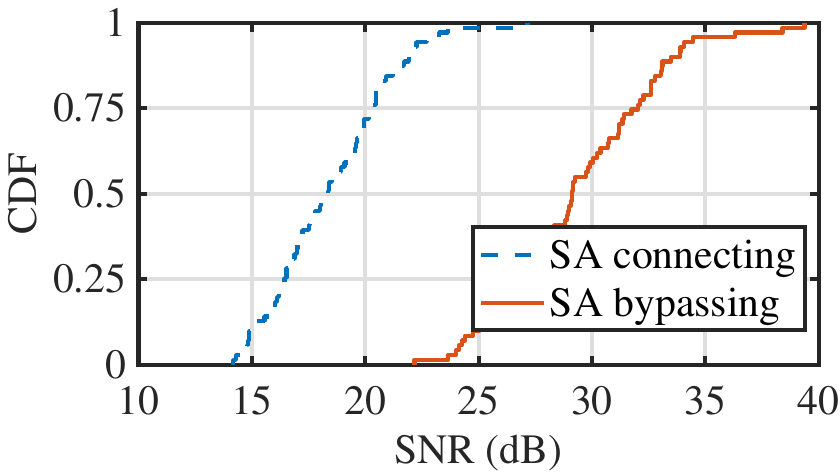}
		\label{sfig:snr_w_wo_sep}
		\vspace{-2.5ex}
	\end{minipage}
}
\subfloat[Throughput.]{ 
	\begin{minipage}[b]{.49\linewidth}
		\centering
		\includegraphics[width = \textwidth]{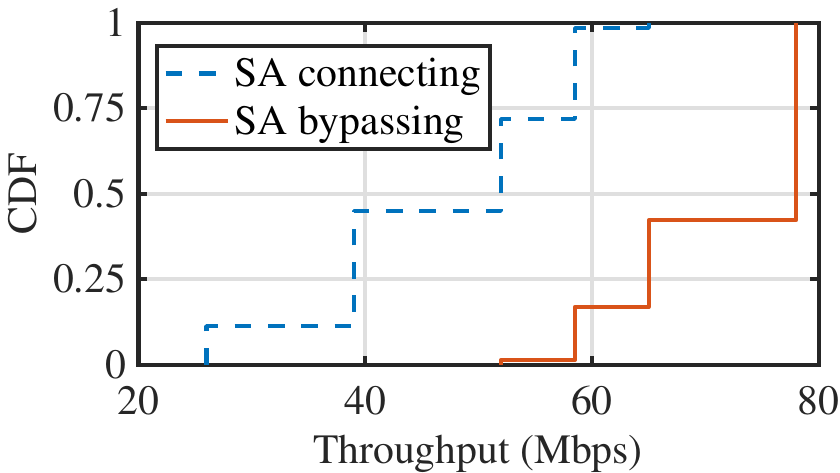}
		\label{sfig:thru_w_wo_sep}
		\vspace{-2.5ex}
	\end{minipage}
}
\caption{The self-adapted Tx-Rx separator (SA) heavily degrades normal Wi-Fi packet reception quality in terms of both (a) SNR and (b) throughput.}
\label{fig:sep_distortion}
\vspace{-1ex}
\end{figure}

Though the successfully implemented Tx-Rx separator can readily enable monostatic sensing on virtually any Wi-Fi NICs, it is not compatible with existing Wi-Fi protocol framework. In particular, as the separator treats all incoming signals indiscriminately, it could strongly affect normal communications due to its filtering (thus distortion) on Rx signals that potentially affects the demodulation performance. \rev{As shown in Fig.~\ref{fig:sep_distortion}, applying the Tx-Rx separator during normal receptions significantly reduces over 12\!~dB SNR and in turn 15\!~Mbps throughput.} Also, it is a waste of computing resource to apply the Tx-Rx separator on normal Rx signals. Note that this problem exists only in the full version of \name, as the partial version use a standalone module to contain the separator that produces monostatic sensing signals. 

For the full version of \name (and its future integrated implementation as an ISAC-ready NIC), we propose and implement the following minor yet critical revision to the protocol. As shown in Fig.~\ref{fig:high_level_arch}, we add a control path (represented by the dotted lines) driven by standard Wi-Fi protocol messages; this entails a function call to the digital cancellator and a hardware interrupt for the analog cancellator. In particular, an Wi-Fi NIC starts with a \textsf{C}-state (for communications), and the transition to an \textsf{M}-state (for monostatic sensing) is invoked by a DATA\footnote{This may include NDP (Null Data Packet) frame not officially standardized by IEEE 802.11, if sensing is required when no data traffic is available.} message containing any Wi-Fi traffic or an ACK message responding to certain data receptions. A transition back from the \textsf{M}-state to the \textsf{C}-state is controlled by a timer fine-tunable to suit surrounding environments. One may also consider a transition from the \textsf{C}-state to the \textsf{B}-state (for bistatic sensing) invoked by a reception of either ACK or DATA from another Wi-Fi NIC, but this is already implicitly assumed by existing Wi-Fi sensing proposals and requires no particular modification to Wi-Fi protocols.

\subsection{Monostatic Channel Feature Estimation} \label{ssec:monochannel}
Different from existing Wi-Fi sensing proposals hacking Wi-Fi NICs for pure sensing purposes, \name should not operate in such a brute force manner, as it promises to stay compatible with existing Wi-Fi standard. Consequently, the sensing information that piggybacks on data packets (for both monostatic and bistatic) often arrives irregularly due to the inherent nature of Wi-Fi data traffic, rendering existing channel feature estimation techniques largely invalid. 

\begin{figure}[t]
\setlength\abovecaptionskip{6pt}
\vspace{-2.5ex}
\raggedleft
\!\!\subfloat[Regular packets.]{ 
	\begin{minipage}[b]{.49\linewidth}
		\centering
		\includegraphics[width = \textwidth]{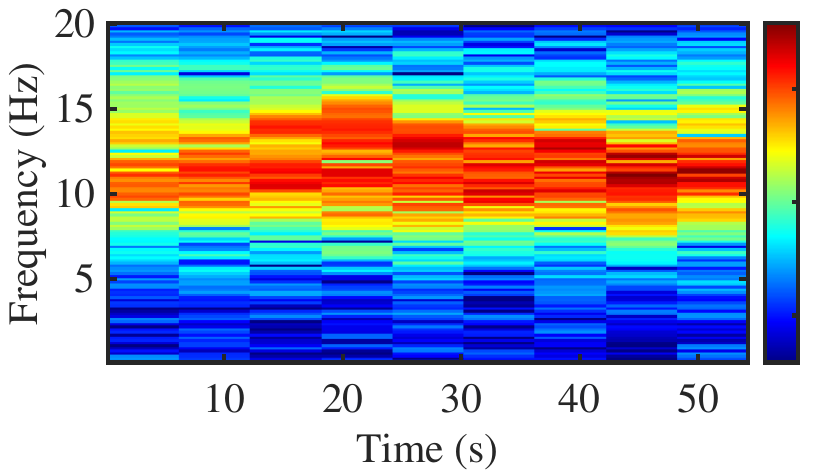}
		\label{sfig:uniform_stft}
		\vspace{-2.5ex}
	\end{minipage}
}
\subfloat[Irregular packets.]{ 
	\begin{minipage}[b]{.49\linewidth}
		\centering
		\includegraphics[width = \textwidth]{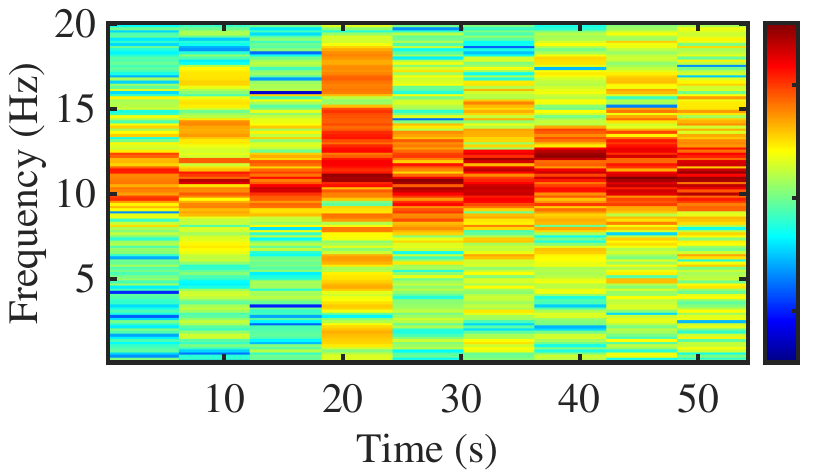}
		\label{sfig:nonuniform_stft}
		\vspace{-2.5ex}
	\end{minipage}
}
\caption{The STFT heatmaps of human slowly walking under (a) regular and (b) irregular packets.}
\label{fig:irregular_packets}
\vspace{-1.5ex}
\end{figure}

We give an instance of human slowly walking indoors to illustrate how irregular packets heavily affect channel feature estimation in Fig.~\!\!\ref{fig:irregular_packets}. 
%
Conventionally, the motion-induced delay $\tau_p^{\mathrm{D}}(t)$ can be estimated using STFT (Short-Time Fourier Transform). Given sensing information conveyed by regular packets, STFT works well to achieve the heatmap of $\tau_p^{\mathrm{D}}(t)$ shown in Fig.~\!\ref{sfig:uniform_stft}: its red parts indicate a high energy concentration ranging from 9 to 15\!~Hz.
However, when applying STFT to sensing information conveyed by irregular packets in practice,
the resulting heatmap of $\tau_p^{\mathrm{D}}(t)$ becomes Fig.~\!\ref{sfig:nonuniform_stft}: the high-energy parts scatter from 0 to 20\!~Hz. In short, irregular packets introduce noises and thus large errors to machine learning classifiers for human activity recognition. To tackle this challenge, we leverage NFFT (Non-uniform Fast Fourier Transform) and sparse optimization to estimate channel features. 

Though the total number of reflections shown in Eqn.~\eqref{eq:rf_csi} can be large, a few reflections should dominate the rest: only reflections with very significant differences in their delays can be identified under a certain bandwidth. Therefore, the path set is sparse and constrained by delayed versions of the know baseband $s(t)$.
%
%
Let the Tx times of the irregular packets be ${\mathbf{T}}^{\mathrm{tx}} = \left[ T_{1}^{\mathrm{tx}}, T_2^{\mathrm{tx}}, \cdots, T_\ell^{\mathrm{tx}}, \cdots  \right]$,
%
%
the vector ${\bm{\Gamma}} = \left[ \tau_p ({\mathbf{T}}_{\ell}^{\mathrm{tx}}) \right]'$ denote the channel features to be estimated, and $\mathcal{F}^{-1}(\bm{\Gamma})$ represent the inverse-NFFT of matrix $\left[ e^{-j2\pi (f_{\mathrm{c}}+k\Delta f) \tau_p({\mathrm{T}}_{\ell}^{\mathrm{tx}})} \right]_{k,\ell}$, then the sparse optimization problem 
can be formulated as:
\begin{align}  \label{eq:opt}
\min \ & \|  { \bm{ \Gamma} } \|_1 \\
\!\!\!\mathrm{s.t.}~~\bigg\Vert\!\sum_p\!x\!\left(n T_{\mathrm{s}}, \tau_{p > 0} ({\mathbf{T}}^{\mathrm{tx}})\right) & -  \mathcal{F}^{-1} ({\bm{\Gamma}}) \cdot s\!\left(n T_{\mathrm{s}} +{\mathbf{T}}^{\mathrm{tx}}\right)\!\bigg\Vert _2^2 = 0 \nonumber
\end{align}
where $\|\cdot\|_1 $ and $\|\cdot\|_2 $ refer to $L^1$ and $L^2$ norms, respectively. We adopt ADMM~\cite{ADMM-FTML11} to solve this problem.

\subsection{Collaborative MIMO Sensing} \label{ssec:collaborate}
With every Wi-Fi NIC equipped with the monostatic (thus standalone) sensing capability, a large-scale ISAC system with a much wider coverage and operating on both monostatic and bistatic modes can be established, by coordinating a set of widely deployed Wi-Fi APs via the Internet. However, the underlying coordination, sitting at the distributed system level, is far beyond the scope of our paper; we hence leave it as a direction for future exploration. In the following, we consider a small set of Wi-Fi NICs co-existing in the same collision domain (with two communicating parties as a special case), and we discuss how to coordinate them in order to seamlessly leverage their monostatic and bistatic sensing capabilities. \rev{It is noted that due to the CSMA/CA mechanism adopted by the Wi-Fi networks, only one Tx-Rx pair is allowed to
communicate at a time slot, and there is nearly zero-interference among multiple Wi-Fi devices.}

Assuming that Wi-Fi devices within the same collision domain are aware of each other in terms of IDs (MAC addresses) and physical locations,\footnote{This is a necessary yet reasonable assumption, as sensing information would become meaningless without these baselines and a database containing such information can be preset when deploying each Wi-Fi device.} our collaborative MIMO sensing scheme demands every of them to periodically share their sensing information using broadcast. Here the sensing information may refer to either individual estimation results or (compressed) raw CSI data. After receiving a sufficient amount of shared sensing information, each device invokes a fusion algorithm to combine these information into a final estimation result. As we are after a readily deployable fusion method to achieve this goal, a maximum likelihood algorithm popular in the radar community is adopted~\cite{3D-UWB-RadarConf}.

\section{Implementation \& Benchmarking} \label{sec:bench}
After elaborating on the implementation details, we evaluate the basic functions of \name in this section.

\subsection{Implementation and Experiment Setup} \label{ssec:impl}
\begin{figure}[b]
\setlength\abovecaptionskip{6pt}
\vspace{-2ex}
\centering
\subfloat[Full version.]{ 
\begin{minipage}[b]{.50\linewidth}
	\centering
	\includegraphics[width = \textwidth]{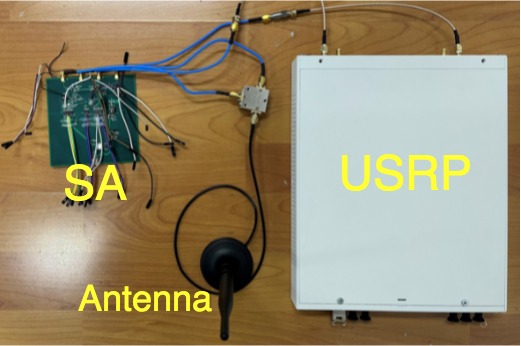}
	\label{sfig:usrp_reduced}
	\vspace{-2.5ex}
\end{minipage}
}
\hfill
\subfloat[Partial version.]{ 
\begin{minipage}[b]{.44\linewidth}
	\centering
	\includegraphics[width = \textwidth]{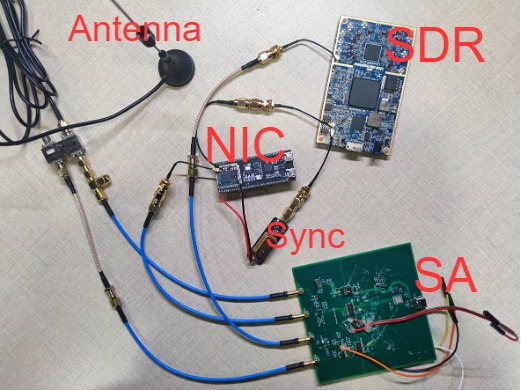}
	\label{sfig:patial2}
	\vspace{-2.5ex}
\end{minipage}
}
\caption{The two prototypes of ISAC-Fi.}
\label{fig:2-photo}
\vspace{-.5ex}
\end{figure}

We construct our own circuit board for the analog cancellator; it applies to both the full and partial \name shown in Fig.~\!\ref{fig:2-photo}. We also implement the digital cancellator, control schedules, as well as various sensing algorithms
in an SDR (Software-Defined Radio) supported by a host PC. The SDR refers to USRP X310~\cite{USRP_x310} and LimeSDR~\cite{limesdr} for the full and partial versions, respectively. \rev{For MIMO configuration, the USRP equips with multiple Tx-Rx separators, and each Tx-Rx separator has one antenna.}
All experiments are done under two scenarios with irregular packets and also other background Wi-Fi traffics: 1) media streaming (UCF101~\cite{UCF101}) and 2) online gaming (StarCraft~\cite{starcraft}).
%

\paragraph{Tx-Rx Separator and Common Procedures} 
A circulator, CentricRF CF2040~\cite{RFcirculator}, or a hybrid coupler, TTM X4C25L1-03G~\cite{RFcoupler}, is employed to respectively suit the full or partial version. 
To process the output of the circulator or hybrid coupler, the analog cancellator is designed as a $5 \times 5 \!~\mathrm{cm}^2$ Printed Circuit Board made in FR-4. We adopt LTC5589~\cite{DQMltc5589} as the DQM; it enables direct modulation of IQ baseband signals at 2.4\!~GHz carrier frequency and its Serial Peripheral Interface can be used to control the Tx gain, supply current, phase imbalance, etc. While the 
digital cancellator is run by the SDR, we further design a General Purpose Input/Output board based on STM32 (an ARM-based MCU) to control the self-adapted Tx-Rx separation involving an RF switch HMC545A~\cite{RFswitch}; the STM32 is in turn controlled by a host PC via USB.
Monostatic sensing algorithms handling irregular packets (as presented in Sec.~\ref{ssec:monochannel}) are implemented in PC, and we further realize the collaborative sensing based on 
MQTT~\cite{mqtt}, a lightweight publish/subscribe messaging protocol for remote devices information exchange.
%
%

\paragraph{Digital Processing and Control Protocols} 
%
For the full version,  we implement the whole Wi-Fi OFDM PHY supporting 20\!~MHz bandwidth, constellations from BPSK to 64 QAM, and all channel codes (with 1/2, 2/3, 3/4, and 5/6 coding rate).  
To let monostatic sensing compatible with CSMA/CA, \name stays normal (the \textsf{C}-state defined in Sec.~\ref{ssec:protocol}), and leverages the Received Signal Strength Indicator to determine whether a channel is idle. When transmitting data packets, the monostatic sensing (the \textsf{M}-state) is invoked to enable the Tx-Rx separator; the transition back to the \textsf{C}-state is triggered by a timer, or the completion of transmission (maximum Wi-Fi frame duration 5.484\!~ms~\cite{wifi80211}), whichever is sooner.
Bistatic sensing (the \textsf{B}-state) is triggered by packet receptions from another Wi-Fi NIC, and the transition back to the \textsf{C}-state naturally follows the completion of reception.
%
%

For the partial version, LimeSDR acts as the sensing module, while ESP32~\cite{ESP32} (an ARM-based MCU with integrated Wi-Fi) is chosen as the Wi-Fi NIC, which already offers the Wi-Fi protocol stack.
%
%
%
To synchronize LimeSDR and ESP32, we design an external 40\!~MHz clock board based on a Temperature Compensated Crystal Oscillator SiT5356~\cite{SiT5356}.
%
%
Most state transitions are the same as the full version except for the trigger for transiting to the \textsf{M}-state: when the host PC demands the Wi-Fi NIC to transmit data packets via hardware USB interrupt, it also invokes LimeSDR to the start the Tx-Rx separator simultaneously. 
%
%
%

\subsection{Tx-Rx Separation Performance}
We hereby study the performance and impact of Tx-Rx separation. We first quantify the interference cancellation ability of different components in the separator. Then we evaluate the impact of Tx-Rx separation on normal Wi-Fi data traffic. 
%

\begin{figure}[t]
\setlength\abovecaptionskip{8pt}
\centering
\includegraphics[width=.92\linewidth]{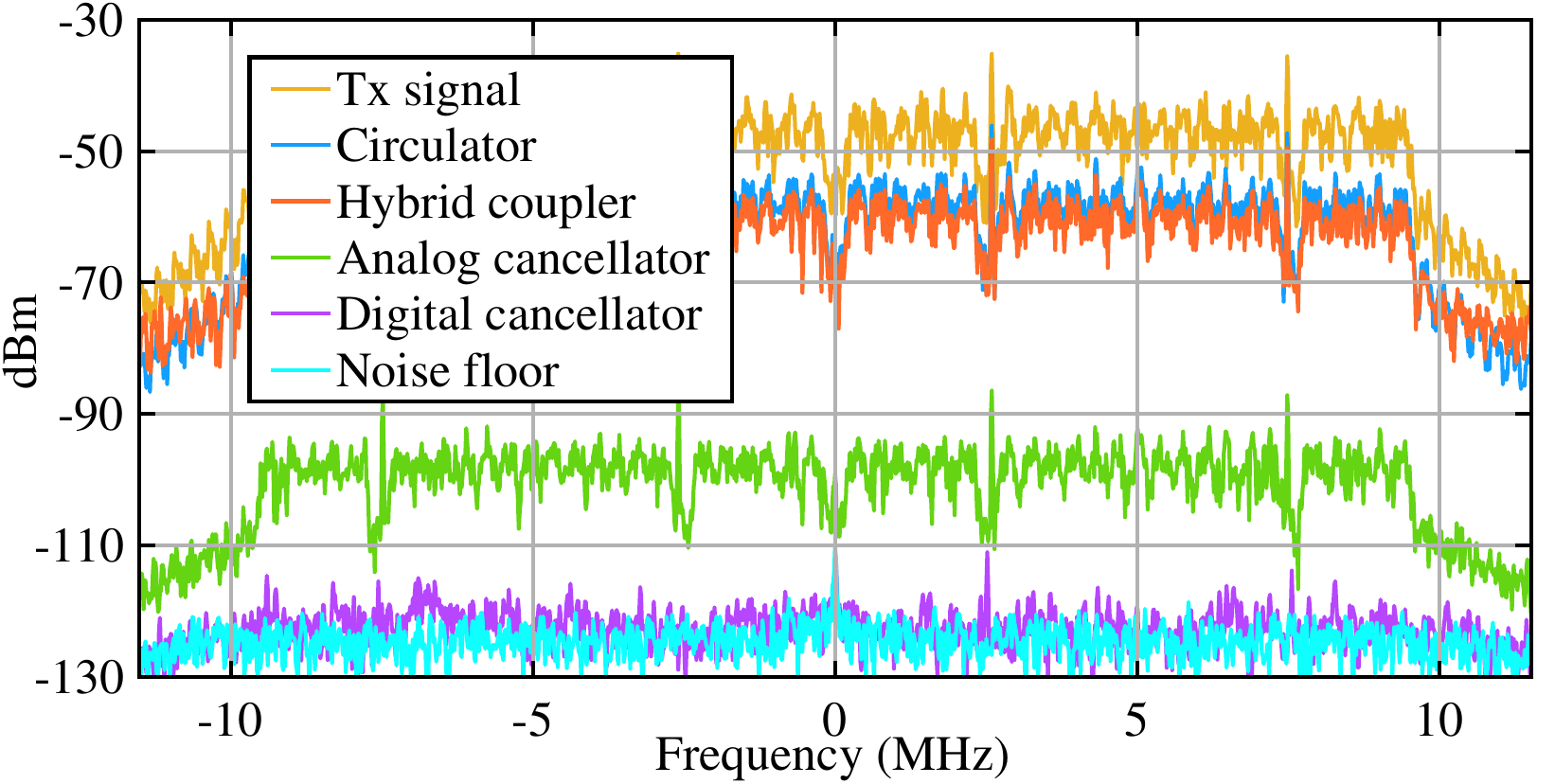}
\caption{Power spectrum of the received baseband signal after various components of Tx-Rx separators.}
\label{fig:tx_rx_sep_eval}
\vspace{-1.5ex}
\end{figure}

We evaluate the performance of Tx-Rx separation under the video streaming scenario with the Tx power set to 5\!~dBm; the results are shown in Fig.~\!\ref{fig:tx_rx_sep_eval}. 
Since the full version with circulator and the partial version with hybrid coupler have nearly the same 12\!~dB cancellation outcome, we only plot the effects of the analog and digital cancellators for the full version. It can be observed that the analog and digital cancellators further reduce the Tx-interference by 40\!~dB and 25\!~dB, respectively. To sum up, the total cancellation is about 77\!~dB, and the power of residue Tx-interference after Tx-Rx separator is very close to the noise floor. 
%

\rev{
\begin{figure}[h]
\setlength\abovecaptionskip{8pt}
\centering
\includegraphics[width=.92\linewidth]{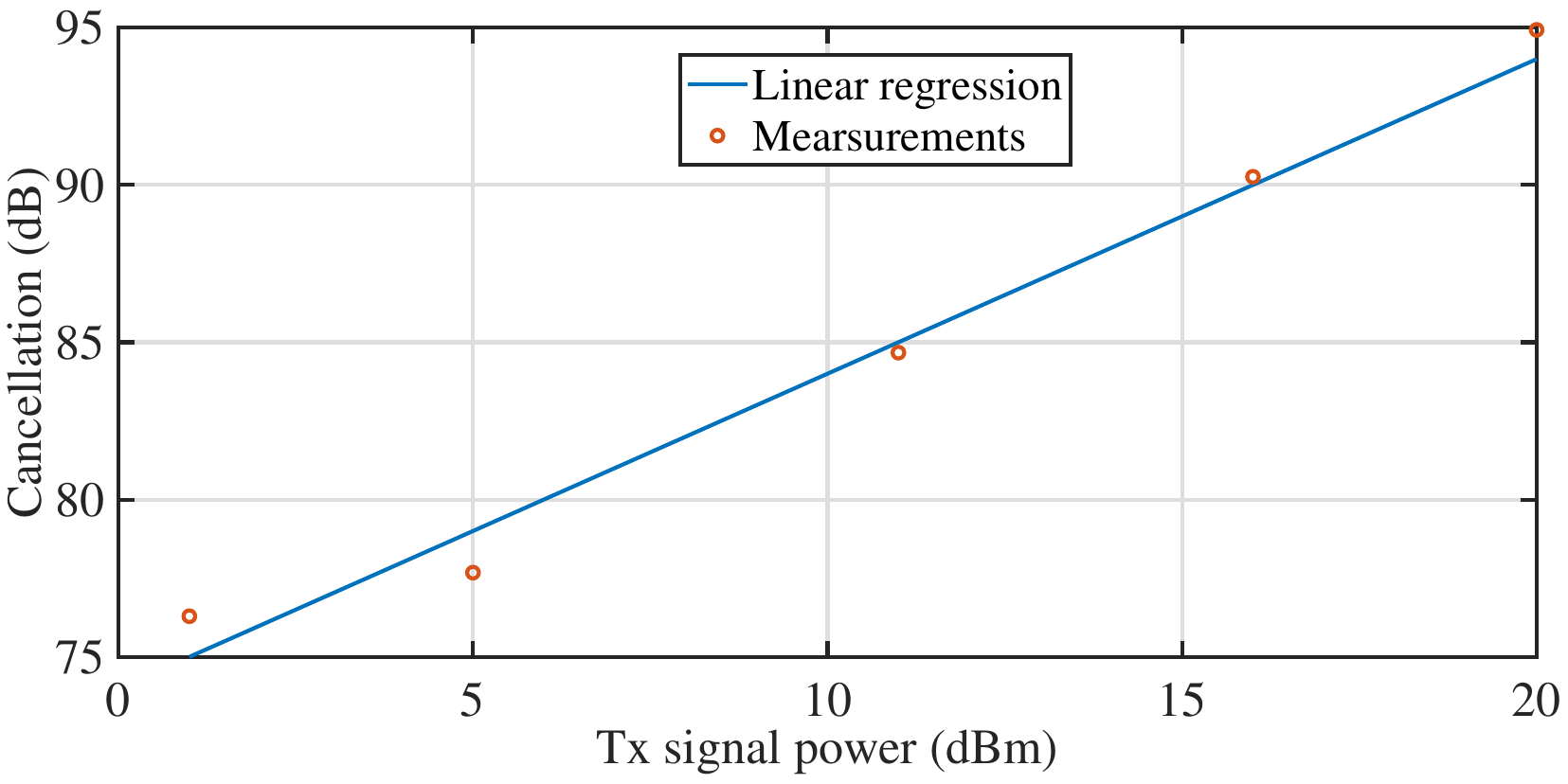}
\caption{The impact of different Tx signal power on the Tx-Rx separator.}
\label{fig:power_tx_sep}
\vspace{-1.5ex}
\end{figure}
We also perform an experiment to study the impact of different Tx signal power on the Tx-Rx separator. We set the parameters of the Tx-Rx separator based on different Tx signal power to keep the power of residue Tx-interference close to noise floor, and plot the cancellation in  Fig.~\ref{fig:power_tx_sep}. Apparently, we can leverage linear regression to model the relationship between Tx power and cancellation, and hence, \name automatically selects parameters for the Tx-Rx separator. Since the residue Tx-interference is close to the noise floor after the Tx-Rx separator, the sensing performance is unaffected by the different Tx power.
}

\begin{figure}[h]
\setlength\abovecaptionskip{6pt}
\vspace{-1.5ex}
\raggedleft
\!\!\subfloat[Packet delay.]{ 
	\begin{minipage}[b]{.49\linewidth}
		\centering
		\includegraphics[width = \textwidth]{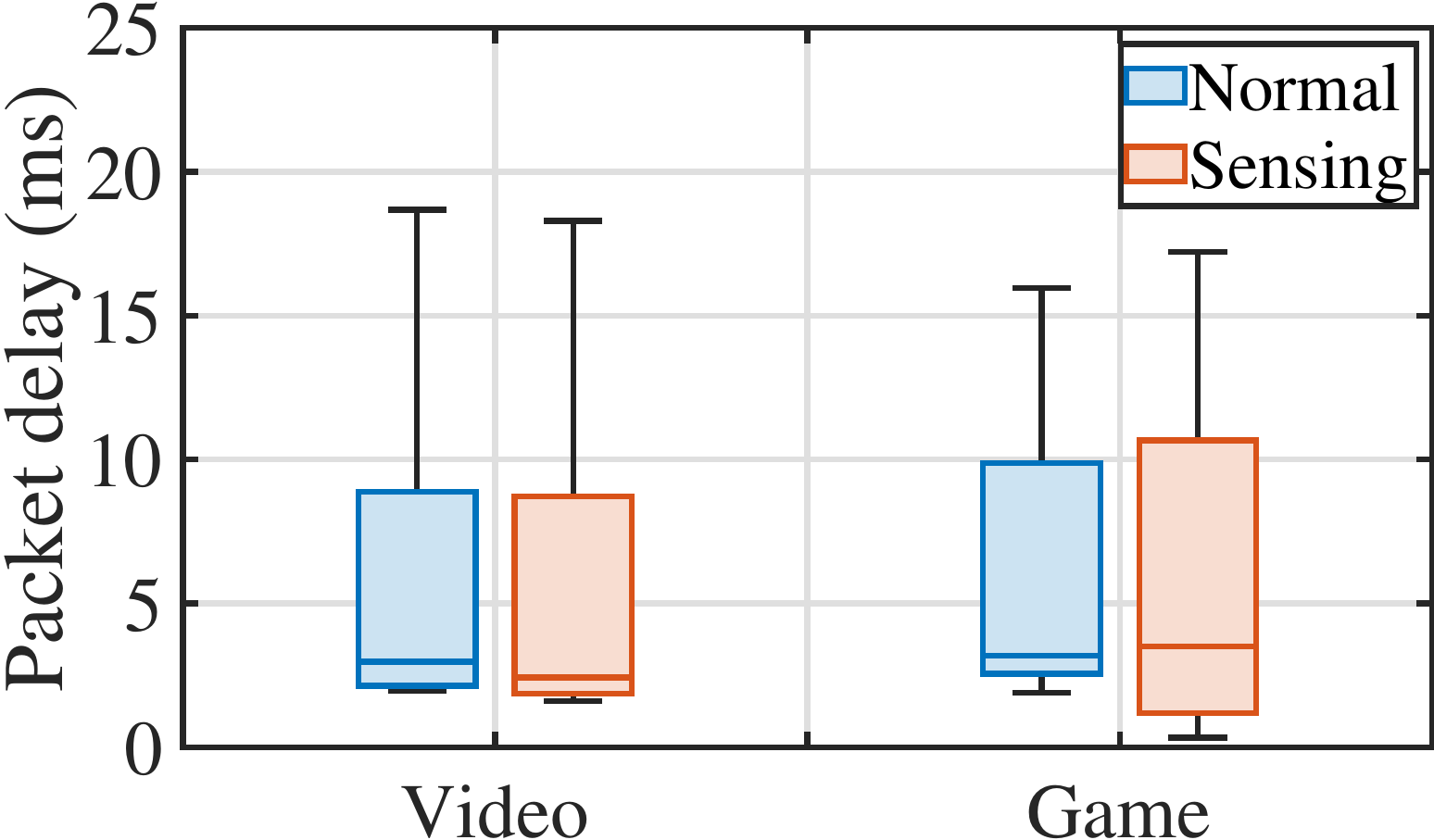}
		\label{sfig:video_game_delay}
		\vspace{-2.5ex}
	\end{minipage}
}
\subfloat[Packet loss rate.]{ 
	\begin{minipage}[b]{.49\linewidth}
		\centering
		\includegraphics[width = \textwidth]{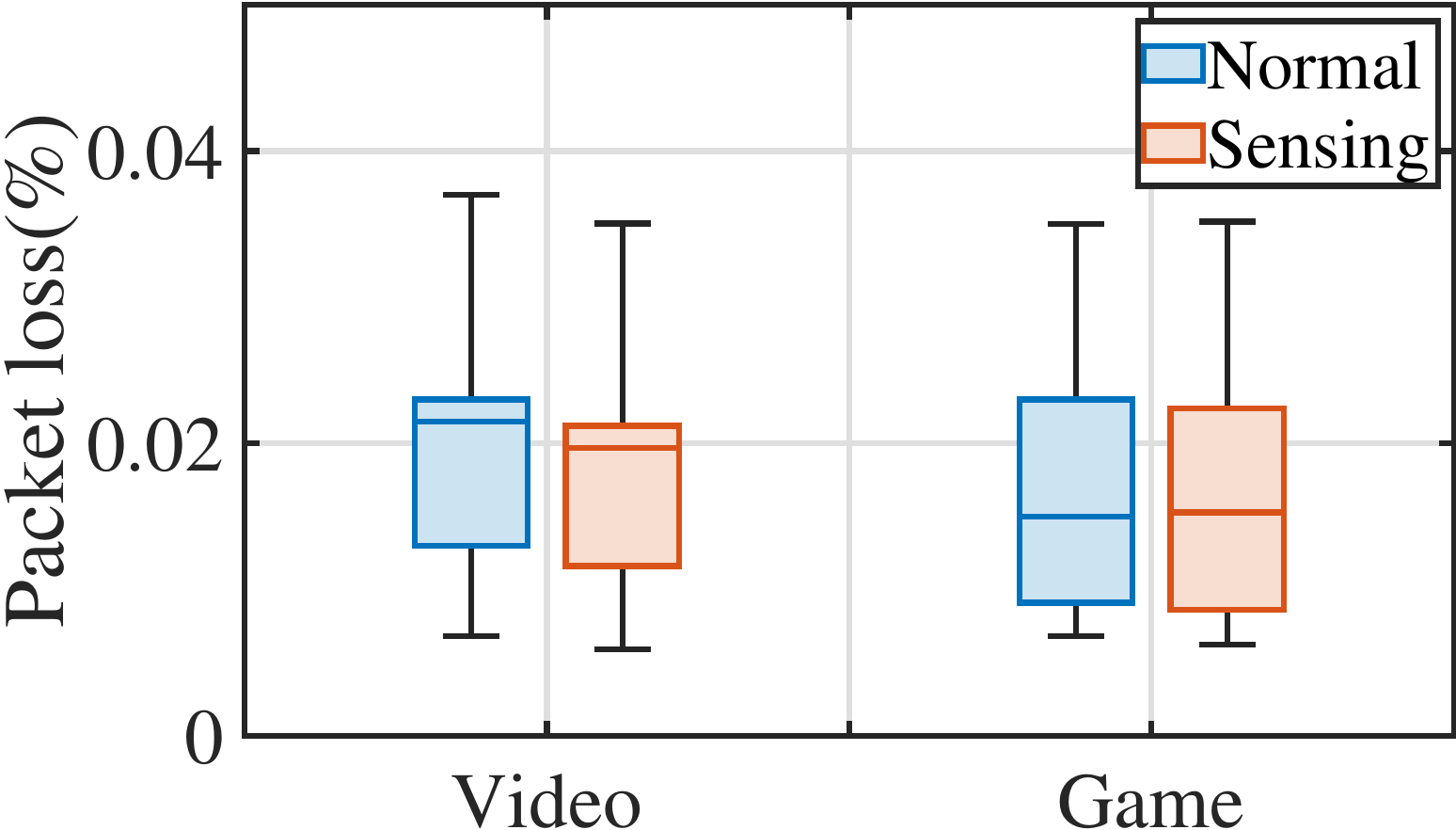}
		\label{sfig:video_game_plr}
		\vspace{-2.5ex}
	\end{minipage}
}
\caption{Impact on normal Wi-Fi communication.}
\label{fig:video_game}
\vspace{-3ex}
\end{figure}

We then study the impact of Tx-Rx separation on Wi-Fi communication, leveraging UDP-based video streaming and online gaming as the testing scenarios since TCP conceals packet loss. Specifically, the USRP-based full version, as it cannot be configured to operate in a multistatic setting, is evaluated by video streaming,
and the partial version is evaluated by both video streaming and online gaming
with at least 3 users/players.
These experiment settings are employed to conduct all remaining experiments. 
Evaluation results of packet delay and packet loss rate are shown in Fig.~\ref{fig:video_game}. The results show that, though \name leverages data packets to perform sensing, it achieves almost the same communication performance as normal Wi-Fi, demonstrating a zero-interference from sensing to communication.
Note that in the experiments, the signals are intentionally attenuated by wall blockage to generate discernible results on packet loss; otherwise they are mostly always 0\%.

\subsection{Ranging Performance} \label{ssec:ranging}
\begin{figure}[t]
\setlength\abovecaptionskip{6pt}
\raggedleft
\vspace{-1.5ex}
\!\!\subfloat[Full version.]{ 
	\begin{minipage}[b]{.49\linewidth}
		\centering
		\includegraphics[width = \textwidth]{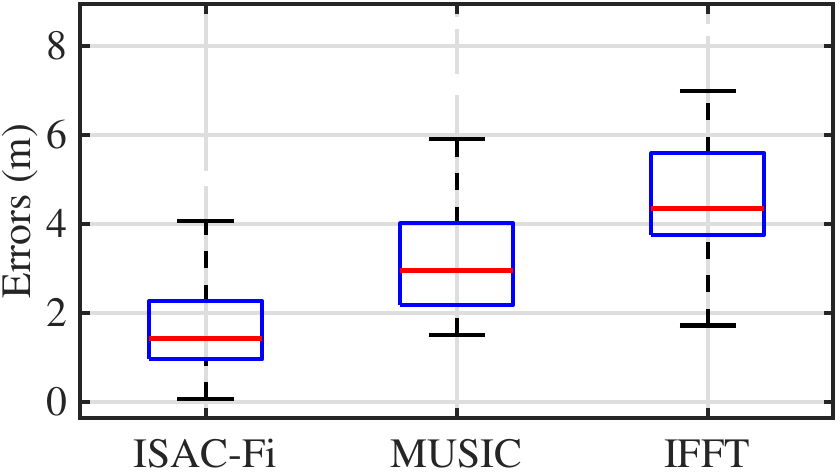}
		\label{sfig:ranging_full}
		\vspace{-2.5ex}
	\end{minipage}
}
\subfloat[Partial version.]{ 
	\begin{minipage}[b]{.49\linewidth}
		\centering
		\includegraphics[width = \textwidth]{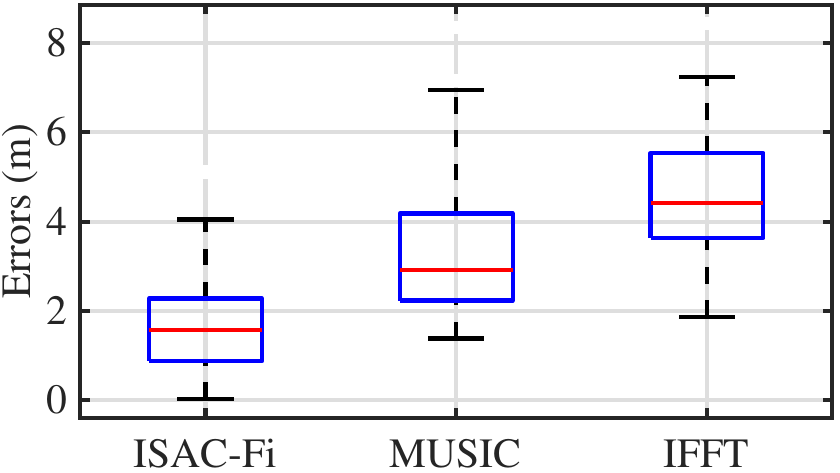}
		\label{sfig:ranging_patial}
		\vspace{-2.5ex}
	\end{minipage}
}
\caption{Ranging error comparisons.}
\label{fig:ranging_errors}
\vspace{-2ex}
\end{figure}

Ranging is a basic yet important function for Wi-Fi sensing, but it can only be achieved under the monostatic mode (as explained in Sec.~\ref{ssec:tfchannel}), whereas existing proposals (e.g., \cite{SpotFi-SIGCOMM15,mDTrack-MobiCom19}) can only perform rough or relative estimations. To demonstrate the ranging performance of \name under irregular data packets using our sensing algorithms introduced in Sec.~\ref{ssec:monochannel}, we choose both IFFT and MUSIC algorithms~\cite{xiong2015tonetrack} as the baselines. In this experiment,  we fix a metal cylinder~(radius 0.1\!~m and height 1.2\!~m) on a robot car. Controlling the car remotely to move from 1\!~m to 15\!~m with a 1\!~m step size in a corridor, we obtain the ranging errors shown in Fig.~\ref{fig:ranging_errors}. Apparently, the same algorithm can obtain similar performance on both full and partial versions.
Also, the medians of ranging errors are 1.42\!~m, 2.84\!~m, and 4.32\!~m for \name, MUSIC, and IFFT, respectively, which can mostly be attributed to \name's adaptation to irregular data packets.
%
%

\subsection{Motion Sensing Performance}
As discussed in Sec.~\ref{ssec:ambiguity-motion}, the direction of each sensed motion is concretely defined for \name: it is the Tx/Rx-target direction. Therefore, we can fully determine the magnitude and bearing of a motion with at least two \name devices, which can never be truly achieved by the bistatic sensing regardless of how many Wi-Fi NICs are involved. To validate the claimed performance of \name, we let the robot car move at different speed range from 0.6\!~m/s to 3.5\!~m/s  in a hall of $20 \times 10\!~\text{m}^2$, and we set a  10\!~m spacing between two \name devices to measure the velocity using monostatic sensing. To obtain ground truth, we let two TI 77\!~GHz millimeter-wave radars~\cite{IWR1443} concurrently perform sensing alongside \name.  

\begin{figure}[b]
\setlength\abovecaptionskip{6pt}
\raggedleft
\vspace{-2ex}
\!\!\subfloat[Full version.]{ 
	\begin{minipage}[b]{.49\linewidth}
		\centering
		\includegraphics[width = \textwidth]{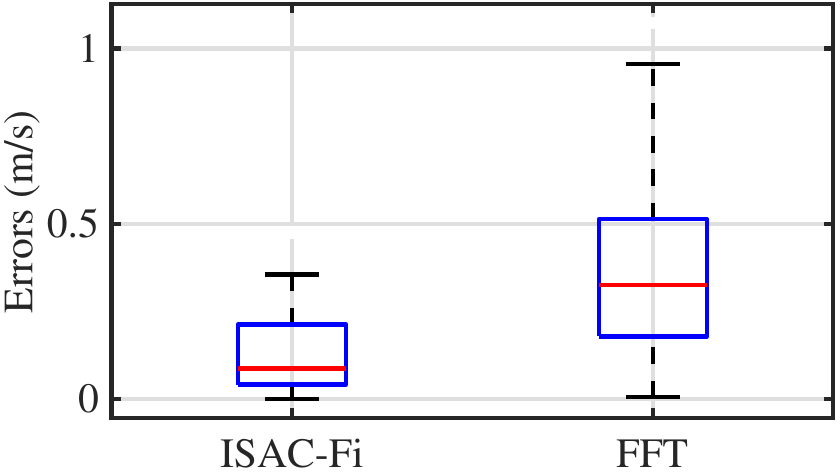}
		\label{sfig:motion_err_full}
		\vspace{-2.5ex}
	\end{minipage}
}
\subfloat[Partial version.]{ 
	\begin{minipage}[b]{.49\linewidth}
		\centering
		\includegraphics[width = \textwidth]{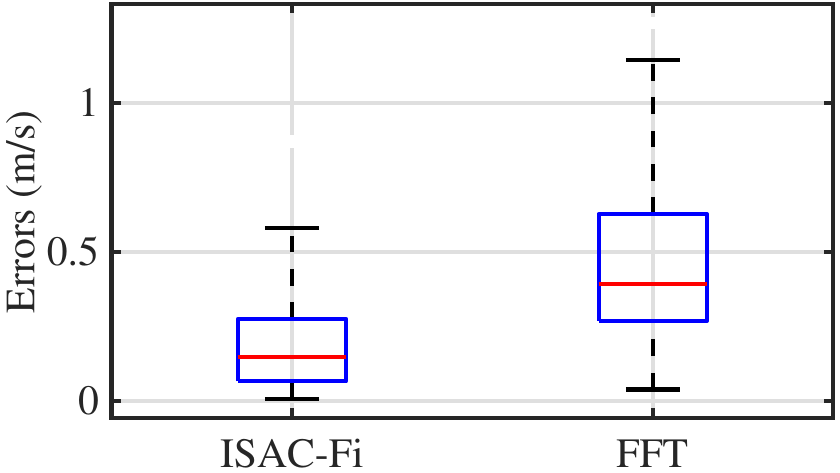}
		\label{sfig:motion_err_partial}
		\vspace{-2.5ex}
	\end{minipage}
}
\caption{The velocity errors of \name and FFT.}
\label{fig:motion_err}
\vspace{-.5ex}
\end{figure}

The evaluation results shown in Fig.~\ref{fig:motion_err} clearly demonstrate that \name (both full and partial versions) achieves much lower estimation errors than the baseline algorithm leveraging FFT~\cite{kleinhempel1993automobile}. However, the partial version seems to perform slightly worse than the full one. Unlike the ranging estimation in Sec.~\ref{ssec:ranging} relying on individual packets, motion sensing depends on a series of timestamped packets. Therefore, the partial version, compared with the full one, the Tx times of the irregular packets may be not exactly counted due to its relatively casual construction: as introduced in Sec.~\ref{ssec:impl}, the triggering signals of the partial version come from the host PC all the way down to the SDR, 
passing through application, OS kernel, driver, and hardware, potentially bringing unpredictable temporal uncertainties. Fortunately, as our partial \name is just a makeshift for backward compatibility, we believe an integrated design for a true \name implementation (emulated by the full version) should not have such constraints.

\section{Evaluation} \label{sec:eval}
Though applications of device-free Wi-Fi sensing are plentiful, they can be roughly classified into three categories, namely localization, activity recognition, and imaging. Therefore, we evaluate \name's performance on these categories, while comparing it with representative proposals for each category whenever applicable. 
However, as \name is meant to introduce a fundamentally new sensing framework rather than focusing on any specific sensing algorithm, 
our evaluations aim to demonstrate the wide capability of \name, in addition to its improvements on representative proposals thanks to the more diversified information brought by \name.

\subsection{Device-free Localization}
As one of the key and novel applications of Wi-Fi sensing, \textit{device-free localization} frees users from holding a Wi-Fi equipped device and solely relies on the deployed Wi-Fi infrastructure to capture the user locations~\cite{LiFS-MobiCom16,mDTrack-MobiCom19}. Nonetheless, existing bistatic sensing proposals fail to totally fulfill the critical demands raised by this challenging application, mainly due to its incompetence in accurately estimating temporal features (as explained in Sec.~\ref{ssec:tfchannel}). Therefore, we choose mD-Track~\cite{mDTrack-MobiCom19} (the latest bistatic sensing proposal on device-free sensing) as the comparison baseline in this section, intending to demonstrate the advantage of \name's monostatic sensing over bistatic sensing. \rev{In addition, the calibration algorithms of bistatic sensing mode are presented in mD-Track~\cite{mDTrack-MobiCom19}.}
%
\begin{figure}[b]
\setlength\abovecaptionskip{8pt}
\vspace{-2ex}
\raggedleft
\!\!\subfloat[Localization errors.]{ 
	\begin{minipage}[b]{.49\linewidth}
		\centering
		\includegraphics[width = \textwidth]{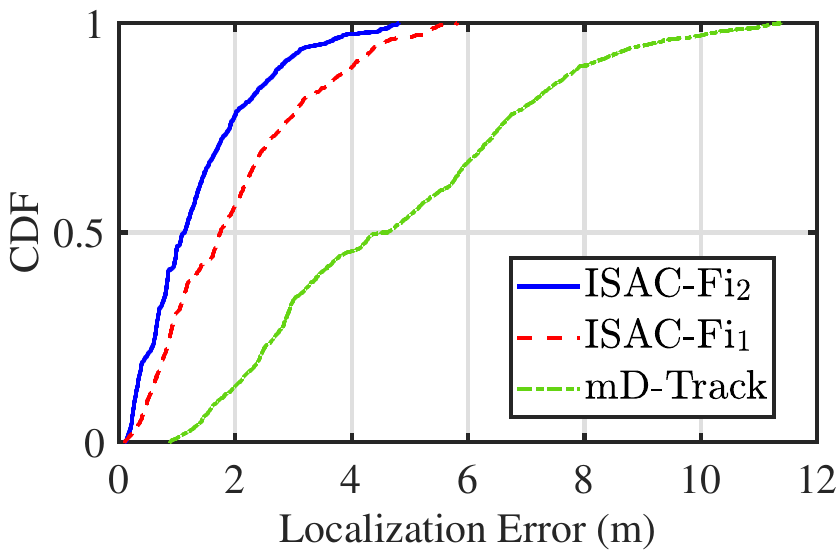}
		\label{sfig:loc}
		\vspace{-2.5ex}
	\end{minipage}
}
\subfloat[ToF errors.]{ 
	\begin{minipage}[b]{.49\linewidth}
		\centering
		\includegraphics[width = \textwidth]{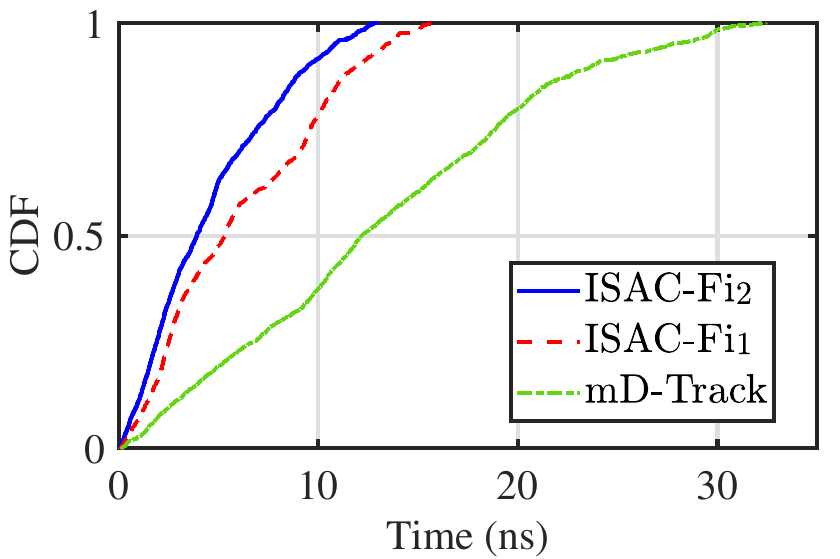}
		\label{sfig:loc2}
		\vspace{-2.5ex}
	\end{minipage}
}
\caption{The performance of non-collaborative and collaborative MIMO localization.}
\label{fig:localizationerror}
\vspace{-.5ex}
\end{figure}

We move the robot car to 16 preset locations in a 100~\!m$^2$ lab space. Both \name and mD-Track operate 3 antennas in 2.4~\!GHz with 20~\!MHz bandwidth,
capturing a $3 \times 3 \times 64$ CSI matrix from each packet preamble. Subsequently, channel features are jointly estimated to derive AoA, ToF, and hence the location. 
For each location, we average 40 measurements to derive an error by comparing with the ground truth, so as to derive 100 such errors with 4,000 measurements. 
As \name excels at ToF estimation, we report the CDFs of both localization and ToF errors in Fig.~\ref{fig:localizationerror}, comparing mD-Tracks with two \name localization schemes: while $\mbox{\name}_1$ leverages both ToF and AoA estimated by a single device to infer location, $\mbox{\name}_2$ exploits two collaborative devices to reach the same goal, while jointly estimating their mutual LoS distance at the same time. 
As expected, \name outperforms mD-Track with a median localization error down to 1.12~\!m as opposed to mD-Track’s 4.57~\!m, and $\mbox{\name}_2$ performs slightly better than $\mbox{\name}_1$ due to the collaborative sensing. These results are consistent with our analysis in Sec.~\ref{ssec:tfchannel} that ToFs of propagation paths cannot be accurately estimated under 
the bistatic mode and that \name is designed to tackle this challenge. The performance of mD-Track shown in Fig.~\ref{sfig:loc} is significantly worse than that reported in~\cite{mDTrack-MobiCom19}, because the Tx-Rx LoS distance was manually measured in~\cite{mDTrack-MobiCom19} and their algorithm does not accommodate irregular packets.
In fact, the performance of \name is also slightly below our expectation, possibly confined by the limited bandwidth.
%

%

\subsection{Human Activity Recognition} \label{ssec:har}
Contact-free \textit{human activity recognition} (HAR) plays a key role in a wide range of real-world applications~\cite{EI-MobiCom18, RF-Net}, and existing Wi-Fi-based HAR solutions directly translate CSI to classification results~\cite{wang2015understanding, EI-MobiCom18, guo2017wifi, shi2017smart}. However, due to the ambiguities of bistatic motion sensing mentioned in Sec.~\ref{ssec:ambiguity-motion}, such translations can be misled and thus resulting in degraded performance. Therefore, we want to demonstrate using experiments that \name's monostatic sensing, albeit relying on only a single Wi-Fi device, may achieve comparable or even better performance than existing bistatic solutions with at least two devices involved.
%

In fact, basic HAR may not be a perfect task for evaluating sensing capabilities, because
conventional bistatic Wi-Fi systems may still yield a high accuracy by overfitting the training/validation data. 
Therefore, we choose a more difficult \textit{cross-domain HAR} task for evaluation, where \textit{cross-domain} means that the environments and human subjects used in training and testing can be different. It is known that Wi-Fi signals carry a substantial amount of environment and subject specific information, so a Wi-Fi HAR method has to resolve this entangled information in order to generalize to new domains. Consequently, we select EI~\cite{EI-MobiCom18} as the comparison baseline given i) its cross-domain capabilities achieved by the novel adversarial learning, and ii) its minimal 1Tx - 2Rx multistatic setup for Wi-Fi sensing.
%

We conduct experiments under several typical indoor settings. Both ISAC-Fi and EI send 40 packets per second for 10 seconds, and 64-subcarrier CSIs are extracted.
%
%
We collect a cross-domain dataset by letting 6 male and 4 female subjects perform 6 activities in 10 rooms with different layouts and sizes (ranging from 6 to 50~\!m$^2$). Without loss of generality, the activities include sitting down, standing up, walking, falling down, bending, and lying down. Each activity is performed 2,500 times, and we obtain a total of 15,000 examples of these activity classes. We employ the same classifier architecture adopted by EI~\cite{EI-MobiCom18}; it includes a 3-layer convolutional network, a domain discriminator, and respective losses to achieve environment and subject independence.
%
%
\begin{figure}[h]
\setlength\abovecaptionskip{6pt}
\vspace{-2ex}
\raggedleft
\!\!\subfloat[ISAC-Fi.]{ 
	\begin{minipage}[b]{.49\linewidth}
		\centering
		\includegraphics[width = \textwidth]{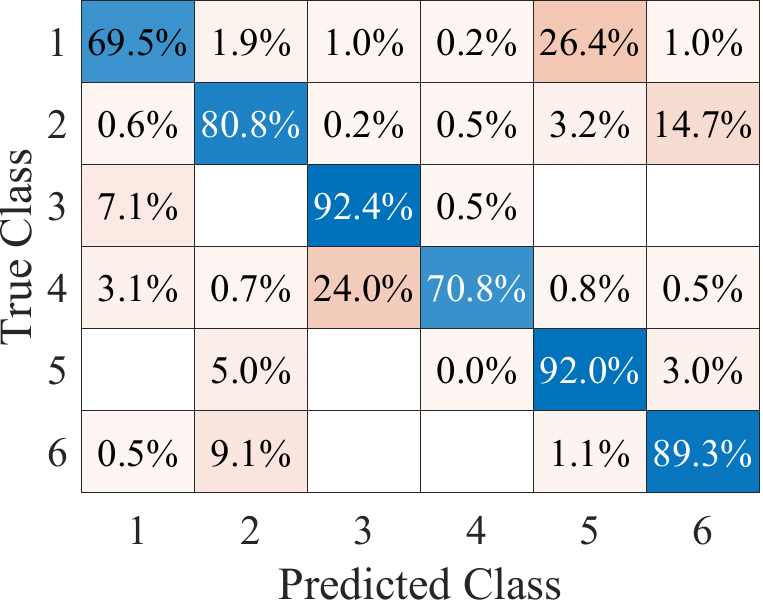}
		\label{sfig:HAR_isac}
		\vspace{-2.5ex}
	\end{minipage}
}
\subfloat[EI (Wi-Fi sensing).]{ 
	\begin{minipage}[b]{.49\linewidth}
		\centering
		\includegraphics[width = \textwidth]{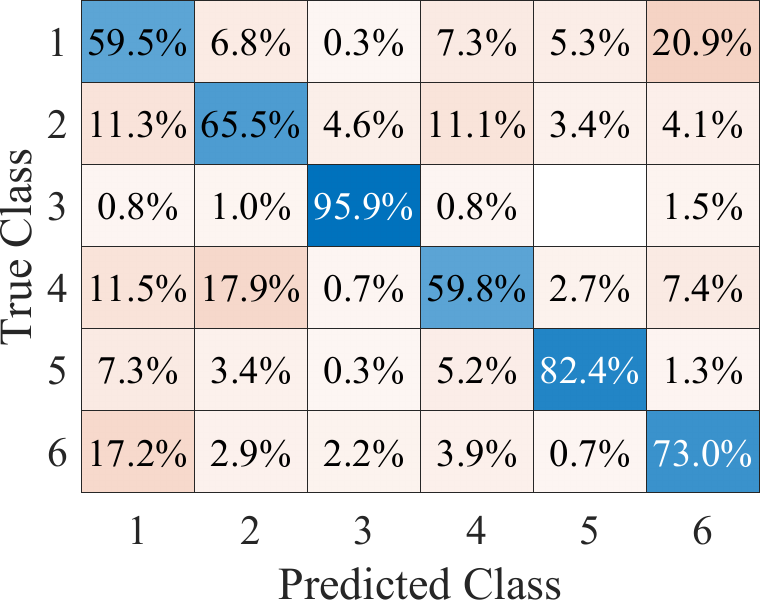}
		\label{sfig:HAR_wifi}
		\vspace{-2.5ex}
	\end{minipage}
}
\caption{Confusion matrices of HAR.}
\label{fig:HAR}
\end{figure}

The evaluation results shown in Fig.~\ref{fig:HAR} indicate that the average HAR accuracy of ISAC-Fi is above 82\%, while that of EI is less than 72\%. The inferior performance of EI can be largely explained by its incompetent cross-domain classification ability, which in turn results from the errors brought by the motion sensing ambiguities typical for a bistatic architecture. The results clearly highlight the efficacy of ISAC-Fi in resolving such ambiguities, allowing it to achieve a higher accuracy in cross-domain HAR.

\subsection{Wi-Fi Imaging} \label{ssec:imaging}
\textit{Wi-Fi imaging} uses Wi-Fi signals for reconstructing images of subjects; it has attracted an increasing attention in recent years~\cite{huang2014feasibility, holl2017holography, karanam20173d, wang2019person, WiPose-MobiCom20}. These proposals aim for generating subject images by leveraging various techniques such as large-scale MIMO~\cite{huang2014feasibility}, synthetic aperture radar~\cite{holl2017holography, karanam20173d}, and multistatic setup~\cite{wang2019person, WiPose-MobiCom20}. Since these proposals often rely on increasing antenna numbers to improve performance, it is almost impossible to quantitatively compare among them. Consequently, we only demonstrate the feasibility of imaging with ISAC-Fi’s novel monostatic sensing in the following. Specifically, we employ the deep learning techniques adopted by~\cite{wang2019person} to translate CSIs captured by ISAC-Fi towards images outlines and skeletons of the subjects.
%

We conduct experiments in rooms and corridors. To enable Wi-Fi imaging, ISAC-Fi leverages its 3-antenna array to improve the spatial diversity in perceiving a subject. We ask human subjects to pose differently at various distances and angles. For each scene, ISAC-Fi averages over 300 packets to obtain a $3 \times 3 \times 150$ CSI matrix, where the first two 3's refer to the antenna number and 150 indicates that 3 packets as a group with only 50 out of 64 subcarriers per packet are used. Meanwhile, a camera next to ISAC-Fi captures a ground truth photo. Since we are interested in outlines and skeletons of human subjects, the photos are further processed to generate binary masks and skeletons of the human subjects for training purposes. We collect a total of 10,000 CSI-image pairs for training deep neural network. 

\begin{figure}[h]
\setlength\abovecaptionskip{8pt}
\vspace{-.5ex}
\centering
\includegraphics[width=.99\linewidth]{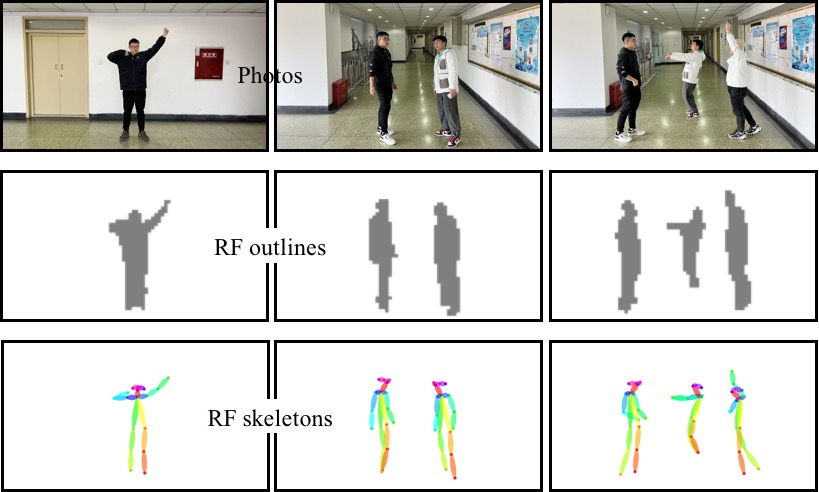}
\caption{Imaging results of human subjects.}
\label{fig:imaging}
\vspace{-.5ex}
\end{figure}


Since all spatial information is embedded in the CSI samples, it is viable to reconstruct human image (outline) and skeleton from its corresponding CSI sample leveraging the deep learning network designed in~\cite{wang2019person}. The network treats the $3 \times 3 \times 150$ CSI samples as $3 \times 3$ images with 150 channels. 
It trains a U-Net~\cite{ronneberger2015u} and skeleton association algorithm~\cite{cao2019openpose} to map the CSIs to outlines and skeletons, leveraging the training data created from ground truth photos. For the training process, we set the batch size to 32, and use the Adam optimizer, whose learning rate and momentum are set to 0.001 and 0.9, respectively.
Fig.~\!\ref{fig:imaging} shows the ground truth photos, RF outlines, and RF skeletons of one, two, and three subjects, respectively. The RF images correctly indicate the number of subjects and clearly show the torso, head, and limbs of each subject, while the skeleton images provide an even sharper characterization of the joint and limb positions. All these results confirm the imaging capabilities of the monostatic sensing adopted by ISAC-Fi. We believe that more realistic imaging results can be achieved if we combine both monostatic and bistatic sensing, but we leave this task to interested researchers.

\section{Related Work and Discussions} \label{sec:survey}
As explained in Sec.~\ref{sec:intro}, Wi-Fi sensing leveraging CSI can be categorized into device-based~\cite{SpotFi-SIGCOMM15,Chronos-NSDI16,DLoc-MobiCom20} and device-free
methods, which can be further divided into three typical applications: localization~\cite{WiDeo-NSDI15,LiFS-MobiCom16,Widar2-MobiSys18,mDTrack-MobiCom19,booranawong2018system,WiVi-SIGCOMM13}, HAR~\cite{WiAG-MobiSys17,EI-MobiCom18,Widar3-MobiSys19,gu2015paws,xiao2021onefi,guo2019wiar} and RF imaging\footnote{This category includes person recognition and/or re-identification~\cite{xu2017radio, avola2022person} that deliver a coarse-grained ``imaging'' as a sub-category.}~\cite{guo2019signal,avola2022human}.
While device-based methods are mainly applied to locate Wi-Fi devices, device-free methods impose no requirement on users but entail a bistatic or even multistatic setting, which often fails to support full-fledged Wi-Fi ISAC systems due to their technology deficiencies such as capable of handling only one person.
Moreover, existing proposals on device-free Wi-Fi sensing have largely remained as experimental prototypes because sensing algorithms are often at odd with Wi-Fi communications. For example, Wi-Vi~\cite{WiVi-SIGCOMM13} applies two Tx antennas to null all static signal/interference and hence totally loses its communication capability. Our \name is specifically proposed and implemented to tackle the challenges faced by existing device-free methods, and it also aims to seamlessly integrate sensing with communication so as to realize the first ISAC-ready Wi-Fi prototype. More importantly, \name can provide more diversified information 
by combining monostatic with bi/multistatic sensing modes.

\rev{Although \name has learned from earlier developments on full-duplex radios (FDR) ~\cite{FDR-MobiCom11,FDR-SIGCOMM13,FDR-MobiCom19}, separating Tx (communication) from Rx (sensing) is fundamentally different from FDR as explained in Sec.~\ref{ssec:separator}. Recent surveys present the architectures, challenges, and opportunities of FDR for future 6G~\cite{smida2023full,alexandropoulos2022full}.} Early proposal WiDeo~\cite{WiDeo-NSDI15} leverages a modified version of FDR to conduct only motion sensing, so it is unable to both locate static subjects and remain compatible with Wi-Fi communications. The closest (title-wise) proposal to our \name in recent literature is \cite{IBFDRC-SysJ20}, yet it merely migrates FDR technique to ISAC scenarios without paying attention to their fundamental differences. Also, the authors in \cite{IBFDRC-SysJ20} never consider the compatibility with Wi-Fi framework and it relies on a proprietary chip for Tx-Rx separation; these have strongly confined its practical feasibility. 
On the contrary, our \name prototypes deploy a critical revision to FDR (see Sec.~\ref{ssec:separator}) and aim to maintain a full compatibility with Wi-Fi communications (see Sec.~\ref{ssec:protocol} and~\ref{ssec:monochannel}), so they are clearly implementable as extended Wi-Fi NICs with necessary manufacturer support. 

It is true that 802.11ax can support 160~\!MHz at 5~\!GHz, yet the whole 160~\!MHz is not always usable due to channel contention. As a result, most of the IoT devices are still leveraging 20~\!MHz bandwidth at 2.4~\!GHz to communicate with each other. Consequently, we start with a basic and common 20~\!MHz bandwidth to design the first full-fledged ISAC system. To the best of our knowledge, our work is pioneering in enabling full-fledged sensing modes over Wi-Fi communication. In other words, we mainly demonstrate design principles in this seminal paper, aiming to deliver guidance for engineering design in future. In this sense, achieving a wider bandwidth (e.g., 160~\!MHz for 802.11ax) should not be the focus of our paper; it is only an engineering extension that can be realized upon our basic design framework by future studies with specific need for it. In fact, a few other challenges still remain for monostatic sensing, such as realizing a large-scale MIMO front-end~\rev{(e.g., millimeter wave~\cite{islam2022integrated})}, and wide-scale distributed MIMO; we also leave these as key directions for our further explorations.


\section{Conclusion} \label{sec:conclusion}
In this paper, we have proposed the idea of making Wi-Fi ISAC-ready and have then reported two prototypes of \name to demonstrate that this idea is completely viable and implementable. We have first motivated our design using several concrete analyses, then we have provided a thorough elaboration on the various aspects of the \name prototypes, followed by extensive evaluations on their performance. Our technical discussions mainly focus on combating (self) Tx-interference and maintaining compatibility with Wi-Fi in terms of both MAC protocol and data traffic aspects. We believe that, by conducting this whole suit of studies, our paper signifies a key step towards a more practical paradigm for future Wi-Fi sensing. In the meantime, we are working towards a full-fledged ISAC-ready design by considering other relevant issues such as expanding frequency bandwidth and accommodating large-scale atenna arrays. 
This work does not raise any ethical issues, as only a few experiment settings involve human subjects (e.g., Sec.~\ref{ssec:separator}, \ref{ssec:har} and~\ref{ssec:imaging}) and they have strictly followed the IRB protocol of our institute. \rev{In the future, we will explore more applications using \name such as vital sign monitoring~\cite{chen2021movi,zheng2020v2ifi}, simultaneous localization and mapping~\cite{wu2023rf}, distributed learning systems~\cite{lin2024split,zhang2024fedac,lin2024efficient,lyu2023optimal,lin2024adaptsfl,yuan2024satsense}, and large language models~\cite{lin2023pushing,fang2024automated,lin2024splitlora}, etc.}



\section*{Acknowledgment}
The work of Jun Luo is partially supported by National Research Foundation (NRF) Future Communications Research \& Development Programme (FCP) grant FCP-NTU-RG-2022-015.


\bibliographystyle{IEEEtran}
\bibliography{reference}

\begin{thebibliography}{10}
\providecommand{\url}[1]{#1}
\csname url@samestyle\endcsname
\providecommand{\newblock}{\relax}
\providecommand{\bibinfo}[2]{#2}
\providecommand{\BIBentrySTDinterwordspacing}{\spaceskip=0pt\relax}
\providecommand{\BIBentryALTinterwordstretchfactor}{4}
\providecommand{\BIBentryALTinterwordspacing}{\spaceskip=\fontdimen2\font plus
\BIBentryALTinterwordstretchfactor\fontdimen3\font minus \fontdimen4\font\relax}
\providecommand{\BIBforeignlanguage}[2]{{%
\expandafter\ifx\csname l@#1\endcsname\relax
\typeout{** WARNING: IEEEtran.bst: No hyphenation pattern has been}%
\typeout{** loaded for the language `#1'. Using the pattern for}%
\typeout{** the default language instead.}%
\else
\language=\csname l@#1\endcsname
\fi
#2}}
\providecommand{\BIBdecl}{\relax}
\BIBdecl

\bibitem{RADAR-INFOCOM00}
P.~Bahl and V.~Padmanabhan, ``{RADAR: An In-Building RF-based User Location and Tracking System},'' in \emph{Proc. of the 19th IEEE INFOCOM}, 2000, pp. 775--784.

\bibitem{Horus-MobiSys05}
M.~Youssef and A.~Agrawala, ``{The Horus WLAN Location Determination System},'' in \emph{Proc. of the 3rd ACM MobiSys}, 2005, p. 205–218.

\bibitem{CSI-CCR11}
D.~Halperin, W.~Hu, A.~Sheth, and D.~Wetherall, ``{Tool Release: Gathering 802.11n Traces with Channel State Information},'' \emph{ACM SIGCOMM Comput. Commun. Rev.}, vol.~41, no.~1, p.~53, 2011.

\bibitem{WiDeo-NSDI15}
K.~Joshi, D.~Bharadia, M.~Kotaru, and S.~Katti, ``{WiDeo: Fine-grained Device-free Motion Tracing using RF Backscatter},'' in \emph{Proc. of the 12th USENIX NSDI}, 2015, pp. 189--204.

\bibitem{LiFS-MobiCom16}
J.~Wang, H.~Jiang, J.~Xiong, K.~Jamieson, X.~Chen, D.~Fang, and B.~Xie, ``{LiFS: Low Human-Effort, Device-Free Localization with Fine-Grained Subcarrier Information},'' in \emph{Proc. of the 22nd ACM MobiCom}, 2016, p. 243–256.

\bibitem{lin2022tracking}
Z.~Lin, L.~Wang, J.~Ding, Y.~Xu, and B.~Tan, ``Tracking and transmission design in terahertz v2i networks,'' \emph{IEEE Transactions on Wireless Communications}, vol.~22, no.~6, pp. 3586--3598, 2022.

\bibitem{Widar2-MobiSys18}
K.~Qian, C.~Wu, Y.~Zhang, G.~Zhang, Z.~Yang, and Y.~Liu, ``{Widar2.0: Passive Human Tracking with a Single Wi-Fi Link},'' in \emph{Proc. of the 16th ACM MobiSys}, 2018, p. 350–361.

\bibitem{lin2022v2i}
Z.~Lin, L.~Wang, J.~Ding, Y.~Xu, and B.~Tan, ``V2i-aided tracking design,'' in \emph{Proc. ICC 2022}, 2022, pp. 291--296.

\bibitem{mDTrack-MobiCom19}
Y.~Xie, J.~Xiong, M.~Li, and K.~Jamieson, ``{mD-Track: Leveraging Multi-Dimensionality for Passive Indoor Wi-Fi Tracking},'' in \emph{Proc. of the 25th ACM MobiCom}, 2019, pp. 8:1--16.

\bibitem{WiVi-SIGCOMM13}
F.~Adib and D.~Katabi, ``{See through Walls with Wi-Fi!}'' in \emph{Proc. of the 27th ACM SIGCOMM}, 2013, p. 75–86.

\bibitem{WiAG-MobiSys17}
A.~Virmani and M.~Shahzad, ``{Position and Orientation Agnostic Gesture Recognition Using WiFi},'' in \emph{Proc. of the 15th ACM MobiSys}, 2017, p. 252–264.

\bibitem{qiu2024ifvit}
Y.~Qiu, H.~Chen, X.~Dong, Z.~Lin, I.~Y. Liao, M.~Tistarelli, and Z.~Jin, ``Ifvit: Interpretable fixed-length representation for fingerprint matching via vision transformer,'' \emph{arXiv preprint arXiv:2404.08237}, 2024.

\bibitem{EI-MobiCom18}
W.~Jiang, C.~Miao, F.~Ma, S.~Yao, Y.~Wang, Y.~Yuan, H.~Xue, C.~Song, X.~Ma, D.~Koutsonikolas, W.~Xu, and L.~Su, ``{Towards Environment Independent Device Free Human Activity Recognition},'' in \emph{Proc. of the 24th ACM MobiCom}, 2018, pp. 289--304.

\bibitem{Widar3-MobiSys19}
Y.~Zheng, Y.~Zhang, K.~Qian, G.~Zhang, Y.~Liu, C.~Wu, and Z.~Yang, ``{Zero-Effort Cross-Domain Gesture Recognition with Wi-Fi},'' in \emph{Proc. of the 17th ACM MobiSys}, 2019, p. 313–325.

\bibitem{RF-Net}
S.~Ding, Z.~Chen, T.~Zheng, and J.~Luo, ``{RF-Net: A Unified Meta-Learning Framework for RF-enabled One-Shot Human Activity Recognition},'' in \emph{Proc. of the 18th ACM SenSys}, 2020, pp. 517--530.

\bibitem{xiao2021onefi}
R.~Xiao, J.~Liu, J.~Han, and K.~Ren, ``{OneFi: One-shot Recognition for Unseen Gesture via COTS WiFi},'' in \emph{Proc. of the 19th ACM SenSys}, 2021, pp. 206--219.

\bibitem{VitalSign-MobiHoc15}
J.~Liu, Y.~Wang, Y.~Chen, J.~Yang, X.~Chen, and J.~Cheng, ``{Tracking Vital Signs During Sleep Leveraging Off-the-Shelf WiFi},'' in \emph{Proc. of the 16th ACM MobiHoc}, 2015, p. 267–276.

\bibitem{Fresnel-UbiComp18}
F.~Zhang, D.~Zhang, J.~Xiong, H.~Wang, K.~Niu, B.~Jin, and Y.~Wang, ``{From Fresnel Diffraction Model to Fine-Grained Human Respiration Sensing with Commodity Wi-Fi Devices},'' in \emph{Proc. of the 20th ACM UbiComp}, 2018, pp. 53:1--23.

\bibitem{Soil-MobiCom19}
J.~Ding and R.~Chandra, ``{Towards Low Cost Soil Sensing Using Wi-Fi},'' in \emph{Proc. of the 25th ACM MobiCom}, 2019, pp. 39:1--16.

\bibitem{Material-MobiCom19}
D.~Zhang, J.~Wang, J.~Jang, J.~Zhang, and S.~Kumar, ``{On the Feasibility of Wi-Fi Based Material Sensing},'' in \emph{Proc. of the 25th ACM MobiCom}, 2019, pp. 41:1--16.

\bibitem{lin2023fedsn}
Z.~Lin, Z.~Chen, Z.~Fang, X.~Chen, X.~Wang, and Y.~Gao, ``Fedsn: A general federated learning framework over leo satellite networks,'' \emph{arXiv preprint arXiv:2311.01483}, 2023.

\bibitem{SpotFi-SIGCOMM15}
M.~Kotaru, K.~Joshi, D.~Bharadia, and S.~Katti, ``{SpotFi: Decimeter Level Localization Using WiFi},'' in \emph{Proc. of 29th ACM SIGCOMM}, 2015, p. 269–282.

\bibitem{Chronos-NSDI16}
D.~Vasisht, S.~Kumar, and D.~Katabi, ``{Decimeter-Level Localization with a Single WiFi Access Point},'' in \emph{Proc. of the 13th USENIX NSDI}, 2016, p. 165–178.

\bibitem{DLoc-MobiCom20}
R.~Ayyalasomayajula, A.~Arun, C.~Wu, S.~Sharma, A.~R. Sethi, D.~Vasisht, and D.~Bharadia, ``{Deep Learning Based Wireless Localization for Indoor Navigation},'' in \emph{Proc. of the 26th ACM MobiCom}, 2020, pp. 17:1--14.

\bibitem{lin2021spatial}
Z.~Lin, L.~Wang, B.~Tan, and X.~Li, ``Spatial-spectral terahertz networks,'' \emph{IEEE Transactions on Wireless Communications}, vol.~21, no.~6, pp. 3881--3892, 2021.

\bibitem{Widar-MobiHoc17}
K.~Qian, C.~Wu, Z.~Yang, Y.~Liu, and K.~Jamieson, ``{Widar: Decimeter-Level Passive Tracking via Velocity Monitoring with Commodity Wi-Fi},'' in \emph{Proc. of the 18th ACM MobiHoc}, 2017, pp. 6:1--10.

\bibitem{WiPose-MobiCom20}
W.~Jiang, H.~Xue, C.~Miao, W.~Shiyang, L.~Sen, C.~Tian, S.~Murali, H.~Hu, Z.~Sun, and L.~Su, ``{Towards 3D Human Pose Construction Using WiFi},'' in \emph{Proc. of the 26th ACM MobiCom}, 2020, pp. 23:1--14.

\bibitem{Kay-Est}
S.~M. Kay, \emph{{Fundamentals of Statistical Signal Processing: Estimation Theory}}.\hskip 1em plus 0.5em minus 0.4em\relax Prentice Hall, 2013.

\bibitem{Octopus-MobiCom21}
Z.~Chen, T.~Zheng, and J.~Luo, ``{Octopus: A Practical and Versatile Wideband MIMO Sensing Platform},'' in \emph{Proc. of the 27th ACM MobiCom}, 2021, pp. 601--614.

\bibitem{FDR-SIGCOMM13}
D.~Bharadia, E.~McMilin, and S.~Katti, ``{Full Duplex Radios},'' in \emph{Proc. of the 27th ACM SIGCOMM}, 2013, pp. 375--386.

\bibitem{USRP_x310}
{Ettus Research}, ``{USRP X310 High Performance Software Defined Radio - Ettus Research},'' \url{https://www.ettus.com/all-products/x310-kit/}, 2014, online; accessed 28 January 2022.

\bibitem{OFDM_TCOM94}
P.~Moose, ``{A Technique for Orthogonal Frequency Division Multiplexing Frequency Offset Correction},'' \emph{IEEE Transactions on Communications}, vol.~42, no.~10, pp. 2908--2914, 1994.

\bibitem{BPath-RADAR03}
R.~Saini, M.~Cherniakov, and V.~Lenive, ``{Direct Path Interference Suppression in Bistatic System: DTV based Radar},'' in \emph{Proc. of IEEE RADAR}, 2003, pp. 309--314.

\bibitem{ADMM-FTML11}
S.~Boyd, N.~Parikh, E.~Chu, B.~Peleato, and J.~Eckstein, ``{Distributed Optimization and Statistical Learning via the Alternating Direction Method of Multipliers},'' \emph{NOW Foundations and Trends in Machine Learning}, vol.~3, no.~1, p. 1–122, 2011.

\bibitem{3D-UWB-RadarConf}
J.~W. Choi and S.~H. Cho, ``{3D Positioning Algorithm Based on Multiple Quasi-Monostatic IR-UWB Radar Sensors},'' in \emph{2017 IEEE RADAR}, 2017, pp. 1531--1535.

\bibitem{limesdr}
{Lime Microsystems Ltd}, ``{LimeSDR},'' \url{https://limemicro.com/products/boards/limesdr/}, 2020, online; accessed 28 January 2022.

\bibitem{UCF101}
{University of Central Florida}, ``{UCF101},'' \url{https://www.crcv.ucf.edu/data/UCF101.php}, 2013, online; accessed 28 January 2022.

\bibitem{starcraft}
{BLIZZARD}, ``{StarCraft II},'' \url{https://starcraft2.com/en-us/}, 2010, online; accessed 28 January 2022.

\bibitem{RFcirculator}
{CentricRF}, ``{CF2040 Circulator 2-4Ghz VSWR 1.35 S Steel SMA 50Watts},'' \url{https://www.centricrf.com/circulators/cf2040-circulator-2-4ghz-sma/}, online; accessed 28 January 2022.

\bibitem{RFcoupler}
{Anaren}, ``{X4C25L1-3G Ultra Low Profile 0603 3dB Hybrid Coupler},'' \url{https://www.ttm.com/en/solutions/rfs-components/xinger-components/90-degree-hybrid-couplers}, 2021, online; accessed 28 January 2022.

\bibitem{DQMltc5589}
{ANALOG DEVICES}, ``{LTC5589 700MHz to 6GHz Low Power Direct Quadrature Modulator},'' \url{https://www.analog.com/en/products/ltc5589.html}, 2016, online; accessed 28 January 2022.

\bibitem{RFswitch}
------, ``{HMC545A GaAs MMIC SPDT Switch},'' \url{https://www.analog.com/media/en/technical-documentation/data-sheets/HMC545A_545AE.pdf}, 2016, online; accessed 28 January 2022.

\bibitem{mqtt}
{MQTT}, ``{MQTT: The Standard for IoT Messaging},'' \url{https://mqtt.org/}, online; accessed 28 January 2022.

\bibitem{wifi80211}
{Matthew S. Gast}, ``{802.11ac: A Survival Guide},'' \url{https://www.oreilly.com/library/view/80211ac-a-survival/9781449357702/ch03.html}, 2013, online; accessed 28 January 2022.

\bibitem{ESP32}
{ESPRESSIF}, ``{ESP32-S2},'' \url{https://www.espressif.com/sites/default/files/documentation/esp32-s2_datasheet_en.pdf}, 2021, online; accessed 28 January 2022.

\bibitem{SiT5356}
{SiTime}, ``{SiT5356},'' \url{https://www.mouser.hk/datasheet/2/371/SiT5356_datasheet-1371850.pdf}, 2021, online; accessed 28 January 2022.

\bibitem{xiong2015tonetrack}
J.~Xiong, K.~Sundaresan, and K.~Jamieson, ``{ToneTrack: Leveraging Frequency-Agile Radios for Time-Based Indoor Wireless Localization},'' in \emph{Proc. of the 21st ACM MobiCom}, 2015, pp. 537--549.

\bibitem{IWR1443}
{Texas Instruments Incorporated}, ``{IWR1843},'' \url{https://www.ti.com/product/IWR1843}, 2019, online; accessed 28 January 2022.

\bibitem{kleinhempel1993automobile}
W.~Kleinhempel, ``Automobile doppler speedometer,'' in \emph{Proc. of VNIS}.\hskip 1em plus 0.5em minus 0.4em\relax IEEE, 1993, pp. 509--512.

\bibitem{wang2015understanding}
W.~Wang, A.~X. Liu, M.~Shahzad, K.~Ling, and S.~Lu, ``{Understanding and Modeling of WiFi Signal based Human Activity Recognition},'' in \emph{Proc. of the 21st ACM MobiCom}, 2015, pp. 65--76.

\bibitem{guo2017wifi}
X.~Guo, B.~Liu, C.~Shi, H.~Liu, Y.~Chen, and M.~C. Chuah, ``{WiFi-enabled Smart Human Dynamics Monitoring},'' in \emph{Proc. of the 15th ACM SenSys}, 2017, pp. 1--13.

\bibitem{shi2017smart}
C.~Shi, J.~Liu, H.~Liu, and Y.~Chen, ``{Smart User Authentication through Actuation of Daily Activities Leveraging WiFi-enabled IoT},'' in \emph{Prof. of the 18th ACM MobiHoc}, 2017, pp. 1--10.

\bibitem{huang2014feasibility}
D.~Huang, R.~Nandakumar, and S.~Gollakota, ``{Feasibility and Limits of Wi-Fi Imaging},'' in \emph{Proc. of the 12th ACM SenSys}, 2014, pp. 266--279.

\bibitem{holl2017holography}
P.~M. Holl and F.~Reinhard, ``{Holography of Wi-Fi Radiation},'' \emph{Physical Review Letters}, vol. 118, no.~18, p. 183901, 2017.

\bibitem{karanam20173d}
C.~R. Karanam and Y.~Mostofi, ``{3D Through-wall Imaging with Unmanned Aerial Vehicles using WiFi},'' in \emph{Proc. of the 16th ACM/IEEE IPSN}, 2017, pp. 131--142.

\bibitem{wang2019person}
F.~Wang, S.~Zhou, S.~Panev, J.~Han, and D.~Huang, ``{Person-in-WiFi: Fine-grained Person Perception using WiFi},'' in \emph{Proc. of the 33rd IEEE ICCV}, 2019, pp. 5452--5461.

\bibitem{ronneberger2015u}
O.~Ronneberger, P.~Fischer, and T.~Brox, ``{U-Net: Convolutional Networks for Biomedical Image Segmentation},'' in \emph{International Conference on Medical Image Computing and Computer-assisted Intervention}.\hskip 1em plus 0.5em minus 0.4em\relax Springer, 2015, pp. 234--241.

\bibitem{cao2019openpose}
Z.~Cao, G.~Hidalgo, T.~Simon, S.-E. Wei, and Y.~Sheikh, ``{OpenPose: Realtime Multi-person 2D Pose Estimation using Part Affinity Fields},'' \emph{IEEE Transactions on Pattern Analysis and Machine Intelligence}, vol.~43, no.~1, pp. 172--186, 2019.

\bibitem{booranawong2018system}
A.~Booranawong, N.~Jindapetch, and H.~Saito, ``{A System for Detection and Tracking of Human Movements using RSSI Signals},'' \emph{IEEE Sensors Journal}, vol.~18, no.~6, pp. 2531--2544, 2018.

\bibitem{gu2015paws}
Y.~Gu, F.~Ren, and J.~Li, ``{PAWS: Passive Human Activity Recognition based on WiFi Ambient Signals},'' \emph{IEEE Internet of Things Journal}, vol.~3, no.~5, pp. 796--805, 2015.

\bibitem{guo2019wiar}
L.~Guo, L.~Wang, C.~Lin, J.~Liu, B.~Lu, J.~Fang, Z.~Liu, Z.~Shan, J.~Yang, and S.~Guo, ``{Wiar: A Public Dataset for WiFi-based Activity Recognition},'' \emph{IEEE Access}, vol.~7, pp. 154\,935--154\,945, 2019.

\bibitem{xu2017radio}
Q.~Xu, Y.~Chen, B.~Wang, and K.~R. Liu, ``{Radio Biometrics: Human Recognition Through a Wall},'' \emph{IEEE Transactions on Information Forensics and Security}, vol.~12, no.~5, pp. 1141--1155, 2017.

\bibitem{avola2022person}
D.~Avola, M.~Cascio, L.~Cinque, A.~Fagioli, and C.~Petrioli, ``{Person Re-identification Through Wi-Fi Extracted Radio Biometric Signatures},'' \emph{IEEE Transactions on Information Forensics and Security}, vol.~17, pp. 1145--1158, 2022.

\bibitem{guo2019signal}
L.~Guo, Z.~Lu, X.~Wen, S.~Zhou, and Z.~Han, ``{From Signal to Image: Capturing Fine-grained Human Poses with Commodity Wi-Fi},'' \emph{IEEE Communications Letters}, vol.~24, no.~4, pp. 802--806, 2019.

\bibitem{avola2022human}
D.~Avola, M.~Cascio, L.~Cinque, A.~Fagioli, and G.~L. Foresti, ``{Human Silhouette and Skeleton Video Synthesis Through Wi-Fi Signals},'' \emph{International Journal of Neural Systems}, vol.~32, p. 2250015, 2022.

\bibitem{FDR-MobiCom11}
M.~Jain, J.~I. Choi, T.~Kim, D.~Bharadia, S.~Seth, K.~Srinivasan, P.~Levis, S.~Katti, and P.~Sinha, ``{Practical, Real-Time, Full Duplex Wireless},'' in \emph{Proc. of the 17th ACM MobiCom}, 2011, p. 301–312.

\bibitem{FDR-MobiCom19}
T.~Chen, M.~Baraani~Dastjerdi, J.~Zhou, H.~Krishnaswamy, and G.~Zussman, ``{Wideband Full-Duplex Wireless via Frequency-Domain Equalization: Design and Experimentation},'' in \emph{Proc. of the 25th ACM MobiCom}, 2019, pp. 3:1--16.

\bibitem{smida2023full}
B.~Smida, A.~Sabharwal, G.~Fodor, G.~C. Alexandropoulos, H.~A. Suraweera, and C.-B. Chae, ``{Full-duplex Wireless for 6G: Progress Brings New Opportunities and Challenges},'' \emph{IEEE Journal on Selected Areas in Communications}, vol.~41, no.~9, pp. 2729--2750, 2023.

\bibitem{alexandropoulos2022full}
G.~C. Alexandropoulos, M.~A. Islam, and B.~Smida, ``{Full-duplex Massive Multiple-Input, Multiple-Output Architectures: Recent Advances, Applications, and Future Directions},'' \emph{IEEE Vehicular Technology Magazine}, vol.~17, no.~4, pp. 83--91, 2022.

\bibitem{IBFDRC-SysJ20}
S.~A. Hassani, V.~Lampu, K.~Parashar, L.~Anttila, A.~Bourdoux, B.~van Liempd, M.~Valkama, F.~Horlin, and S.~Pollin, ``{In-Band Full-Duplex Radar-Communication System},'' \emph{IEEE Systems Journal}, vol.~15, no.~1, pp. 1086--1097, 2020.

\bibitem{islam2022integrated}
M.~A. Islam, G.~C. Alexandropoulos, and B.~Smida, ``{Integrated Sensing and Communication with Millimeter wave Full Duplex Hybrid Beamforming},'' in \emph{Proc. of the IEEE ICC}.\hskip 1em plus 0.5em minus 0.4em\relax IEEE, 2022.

\bibitem{chen2021movi}
Z.~Chen, T.~Zheng, C.~Cai, and J.~Luo, ``{MoVi-Fi: Motion-robust Vital Signs Waveform Recovery via Deep Interpreted RF Sensing},'' in \emph{Proc. of the 27th ACM MobiCom}, 2021.

\bibitem{zheng2020v2ifi}
T.~Zheng, Z.~Chen, C.~Cai, J.~Luo, and X.~Zhang, ``{V2iFi: In-vehicle vital sign monitoring via compact RF sensing},'' \emph{Proc. of the ACM IMWUT}, vol.~4, no.~2, pp. 1--27, 2020.

\bibitem{wu2023rf}
C.~Wu, Z.~Gong, B.~Tao, K.~Tan, Z.~Gu, and Z.-P. Yin, ``Rf-slam: Uhf-rfid based simultaneous tags mapping and robot localization algorithm for smart warehouse position service,'' \emph{IEEE Transactions on Industrial Informatics}, vol.~19, no.~12, pp. 11\,765--11\,775, 2023.

\bibitem{lin2024split}
Z.~Lin, G.~Qu, X.~Chen, and K.~Huang, ``Split learning in 6g edge networks,'' \emph{IEEE Wireless Communications}, 2024.

\bibitem{zhang2024fedac}
Y.~Zhang, H.~Chen, Z.~Lin, Z.~Chen, and J.~Zhao, ``Fedac: A adaptive clustered federated learning framework for heterogeneous data,'' \emph{arXiv preprint arXiv:2403.16460}, 2024.

\bibitem{lin2024efficient}
Z.~Lin, G.~Zhu, Y.~Deng, X.~Chen, Y.~Gao, K.~Huang, and Y.~Fang, ``Efficient parallel split learning over resource-constrained wireless edge networks,'' \emph{IEEE Transactions on Mobile Computing}, 2024.

\bibitem{lyu2023optimal}
S.~Lyu, Z.~Lin, G.~Qu, X.~Chen, X.~Huang, and P.~Li, ``Optimal resource allocation for u-shaped parallel split learning,'' in \emph{2023 IEEE Globecom Workshops (GC Wkshps)}, 2023, pp. 197--202.

\bibitem{lin2024adaptsfl}
Z.~Lin, G.~Qu, W.~Wei, X.~Chen, and K.~K. Leung, ``Adaptsfl: Adaptive split federated learning in resource-constrained edge networks,'' \emph{arXiv preprint arXiv:2403.13101}, 2024.

\bibitem{yuan2024satsense}
H.~Yuan, Z.~Chen, Z.~Lin, J.~Peng, Z.~Fang, Y.~Zhong, Z.~Song, and Y.~Gao, ``Satsense: Multi-satellite collaborative framework for spectrum sensing,'' \emph{arXiv preprint arXiv:2405.15542}, 2024.

\bibitem{lin2023pushing}
Z.~Lin, G.~Qu, Q.~Chen, X.~Chen, Z.~Chen, and K.~Huang, ``Pushing large language models to the 6g edge: Vision, challenges, and opportunities,'' \emph{arXiv preprint arXiv:2309.16739}, 2023.

\bibitem{fang2024automated}
Z.~Fang, Z.~Lin, Z.~Chen, X.~Chen, Y.~Gao, and Y.~Fang, ``Automated federated pipeline for parameter-efficient fine-tuning of large language models,'' \emph{arXiv preprint arXiv:2404.06448}, 2024.

\bibitem{lin2024splitlora}
Z.~Lin, X.~Hu, Y.~Zhang, Z.~Chen, Z.~Fang, X.~Chen, A.~Li, P.~Vepakomma, and Y.~Gao, ``Splitlora: A split parameter-efficient fine-tuning framework for large language models,'' \emph{arXiv preprint arXiv:2407.00952}, 2024.

\end{thebibliography}

\end{document}